\documentstyle[aps,eqsecnum,preprint,epsf]{revtex}
\firstfigfalse

\begin{document}
\tighten
\preprint{DESY 95-191,
{\small E}N{\large S}{\Large L}{\large A}P{\small P}-A-547-95,
TUW-95-19
}
\draft
\title{The dynamics of cosmological perturbations
in thermal $\lambda\phi^4$ theory}
\author{Herbert Nachbagauer\thanks{e-mail: herby@lapphp1.in2p3.fr}
}
\address{{Laboratoire de Physique Th\'eorique}
ENSLAPP
\thanks{URA 14-36 du CNRS, associ\'ee \`a l'E.N.S. de Lyon, et au L.A.P.P.
(IN2P3-CNRS) d'Annecy-le-Vieux.}
\\
B.P. 110, F-74941 Annecy-le-Vieux Cedex, France}
\author{Anton K. Rebhan\thanks{On leave of absence from
        Institut f\"ur Theoretische Physik der
        Technischen Universit\"at Wien until November 1, 1995;
        e-mail: rebhana@x4u2.desy.de}}
\address{DESY, Gruppe Theorie,\\
         Notkestra\ss e 85, D-22603 Hamburg, Germany}
\author{Dominik J. Schwarz
        \thanks{e-mail: dschwarz@itp.phys.ethz.ch; address since October 1,
        1995:
        Theoretische Physik, ETH-H\"onggerberg, CH-8093 Z\"urich}}
\address{Institut f\"ur Theoretische Physik,
         Technische Universit\"at Wien,\\
         Wiedner Hauptstra\ss e 8-10/136,
         A-1040 Wien, Austria}

\date{\today}

\maketitle

\begin{abstract}
Using a recent thermal-field-theory approach to cosmological perturbations, the
exact solutions that were found for collisionless ultrarelativistic matter
are generalized
to include the effects from weak self-interactions
in a $\lambda\phi^4$ model through order $\lambda^{3/2}$.
This includes the effects of a resummation of thermal masses and associated
nonlocal gravitational vertices, thus going far beyond
classical kinetic theory. Explicit solutions for all
the scalar, vector, and tensor modes are obtained
for a radiation-dominated Einstein-de Sitter model
containing a weakly interacting
scalar plasma with or without the admixture
of an independent component of perfect radiation fluid.
\end{abstract}



\section{Introduction}

In the standard model of cosmology, the early universe is described
by a homogeneous and isotropic Friedmann-Lema\^{\i}tre-Robertson-Walker
model. Small linear metric perturbations are responsible
for both, the large-scale structure of the present-day
universe and the tiny deviations from the anisotropy of the
cosmic microwave background. The corresponding linear perturbation
theory has been developed by Lifshitz \cite{Lifshitz} half a century ago
who found the solutions when the energy-momentum tensor is
that of a perfect fluid. (For recent reviews of further developments
since see Ref.~\cite{Kodama}
and \cite{Mukhanov}.)
With more complicated forms of matter
it is typically necessary to resort to numerical integrations
\cite{Peebles70,McCone70,Peebles73,Bond83} of
coupled Einstein-Boltzmann equations
\cite{Ehlers,Stewart72}.
In the case of collisionless matter, some analytic results have
been obtained by Zakharov and Vishniak \cite{Zakharov}, but
also there, the perturbation equations were eventually solved
numerically \cite{McCone70,Bond83}.

In Ref.~\cite{Kraemmer}, a novel framework for the study of
cosmological perturbations has been developed which
is based on thermal field theory. In this formalism the
connection between the perturbed metric and the perturbed
energy-momentum tensor is provided by the (thermal) gravitational
polarization tensor. Concentrating on post-Planckian and
post-inflationary epochs, we assume that $T\ll m_{\rm Planck}$ and
therefore that the gravitational field
can be treated as a classical gravitational
background field.
The momentum scale of cosmological perturbations is set by the inverse Hubble
radius $H^{-1}\sim T^2/m_{\rm Planck}$, which is thus much smaller
than the temperature. If the particles are furthermore
ultrarelativistic, i.e., their masses are negligible when compared
to temperature, it is only the high-temperature
limit of the gravitational polarization tensor which
is needed to determine the response of the primordial plasma
to metric perturbations.
The leading high-temperature contributions
to the gravitational polarization tensor, which have been calculated first
in Ref.~\cite{Rebhan91} (see also Ref.~\cite{ABFT}),
describe collisionless ultrarelativistic matter.
Using them to provide the right-hand side of the perturbed Einstein
equations, one
obtains self-consistent and manifestly gauge-invariant perturbation
equations, for which exact, analytic solutions were found in
Ref.~\cite{Kraemmer,Rebhan92a,Schwarz,Rebhan94}.
In this case, one can show \cite{Rebhan94} that
the perturbation equations are equivalent to a certain gauge-invariant
reformulation of the Einstein-Vlasov equations \cite{Kasai}.

In Ref.~\cite{NRS1} we have started to extend the thermal-field-theory
approach to weakly self-interacting thermal matter, for which
we have chosen scalar particles with quartic self-coupling.
With the slight generalization to an $O(N)$-symmetric model,
the Lagrangian is given by
\begin{equation}
\label{L}
{\cal L}(x)=\sqrt{-g(x)}\left\{\textstyle{1\over2}
g^{\mu\nu}\partial_\mu\phi \partial_\nu\phi -
\textstyle{1\over2}\xi R \phi^2 - {3\over (N+2)}\lambda \phi^4 \right\}.
\end{equation}
With $\xi = 1/6$ this Lagrangian is conformally invariant. The precise
value of $\xi$ does not enter in our calculations, since curvature
corrections to the energy-momentum tensor and its perturbations
are suppressed by a factor $T^2/m^2_{\rm Planck}\ll1$. We assume that
self-interactions are much mor important than those,
i.e. $\lambda \gg T^2/m^2_{\rm Planck}$.

The leading self-interaction effects show up as two-loop corrections
to the gravitational polarization tensor (or, in particle physics
terminology, to the thermal graviton self-energy). As it is the
case with its one-loop high-temperature limit,
it satisfies a conformal Ward identity, which
makes it possible
to use momentum-space techniques to evaluate this nonlocal object
completely by going first to flat space-time and then
transforming to the curved background geometry of the cosmological
model, which in virtually all cases of interest is conformally flat.
The corrections to the
perturbation equations of the
collisionless (one-loop)
case turn out to be such that it is still possible
to solve these equations exactly in terms of rapidly converging
power series. The resulting changes turned out to be perturbative
on scales comparable to or larger than the Hubble horizon, whereas
the large-time behaviour of subhorizon-sized perturbations become
increasingly sensitive to formally higher order effects \cite{NRS1}.

In the present paper, we complete the derivation of the effects
proportional to the scalar self-coupling $\lambda$ and go on to
include the next-to-next-to-leading terms,
which are of order $\lambda^{3/2}$.
To this order, we have calculated the gravitational polarization tensor
in Ref.~\cite{NRS2}, which required the use of a resummed perturbation
theory. Besides the necessity to resum the induced thermal masses
($\sim \sqrt{\lambda} T$)
acquired by soft excitations in the scalar plasma, this also
requires a resummation of nonlocal graviton-scalar vertices.
These effects can be included systematically by our
quantum-field-theoretic framework, while it is unclear how they
could be taken into account in a kinetic-theory approach.

Again, it turned out that the results for the
gravitational polarization-tensor which are proportional to $\lambda^{3/2}$
satisfy the conformal
Ward identity which allows us to transform them to curved space.
Although the results are much more complicated than the previous
ones at order $\lambda^1$,
their structure is again such that the perturbation equations can
be solved exactly.
Since the gravitational polarization tensor also satisfies the Ward
identities required by diffeomorphism invariance, we can continue
to work with manifestly gauge invariant variables for metric perturbations.
Following largely the notation of Bardeen \cite{Bardeen}, our
gauge invariant set-up is laid down in Sect.~2 for the background
geometry of the radiation-dominated spatially flat Einstein-de Sitter model.
In Sect.~3, the solutions for scalar, vector, and tensor perturbations
are constructed after allowing also for an arbitrary admixture of
a perfect radiation fluid, with some of the calculational
details relegated
to the Appendix. As found already in Ref.~\cite{NRS1},
the corrections to the solutions caused by the scalar self-interactions
are perturbative except when the horizon has grown to become much larger
than the wavelength (of a Fourier mode) of the perturbation.
In Sect.~4 we show that the behaviour of the perturbative series can
be greatly improved by rewriting it in terms of a Pad\'e approximant.
This can be tested in the limit $N\to\infty$ of our model (\ref{L}).
Encouraged by these findings, we consider also the late-time behaviour
of our solutions in Sect.~5. Sect.~6 summarizes our results.

\section{Gauge-invariant setup}

We will consider a radiation dominated
Einstein-de Sitter background
\begin{equation}
ds^2 = S^2(\tau) \left( -d\tau^2 + \delta_{ij}dx^i dx^j \right)
\label{ds2}
\end{equation}
with $\tau$ being the conformal time which measures
the size of the horizon in comoving coordinates.
The overall evolution is determined by the cosmic scale
factor $S(\tau )$ which satisfies the Friedmann equation
\begin{equation}
H^2 :=
\left({1\over S^2}{d S \over d \tau}\right)^2 = {8\pi G\over 3} \rho \ .
\end{equation}
The background energy density is related to the pressure by the
equation of state $\rho = 3 P.$

In flat space-time the ring-resummed pressure of the
ultrarelativistic matter described by (\ref{L})
reads \cite{Kapusta}
\begin{equation}
P = N {\pi^2 T^4 \over 90} \left(1 - {15\over8}{\lambda\over\pi^2} +
{15\over 2} \left(\lambda\over\pi^2\right)^{\frac32} + O(\lambda^2) \right) \ .
\end{equation}

In the Einstein-de Sitter background
the same expression holds true, but now
with scale dependent
temperature $T(S) = T(S\negthinspace =\negthinspace 1) S^{-1}$. The existence
of a thermal equilibrium state for relativistic matter
is guaranteed by a conformal timelike Killing vector field $ u^\mu / T$.
The corresponding energy momentum tensor
\begin{equation}
T^{\mu}{}_{\nu} = P (4 u^{\mu}u_{\nu} + \delta^{\mu}_{\nu})
\ , \qquad u_{\mu} = S\delta_\mu^0 \ ,
\end{equation}
is formally that of a perfect fluid and is traceless. This reflects the
conformal symmetry of the effective action for ultrarelativistic plasmas.

We make use of the gauge-invariant metric potentials and matter variables
introduced by Bardeen \cite{Bardeen}. The linear perturbations may be split
into scalar, vector, and tensor parts \cite{Lifshitz}. Since we
work in spatially flat space-time all variables can be decomposed into
plane waves with comoving wave number $k$.
Then the linearized Einstein equations may be written in terms of the
variable
\begin{equation}
\label{xdef}
x := \tau k \ ,
\end{equation}
which  measures the
number of (physical) half wave lengths inside the Hubble radius $H^{-1}$, i.e.
\begin{equation}
{x\over \pi} = H^{-1}\left/ \left(\lambda\over 2\right)\right. \ .
\end{equation}
In what follows it will act as both, a dimensionless time variable and
a dimensionless measure of the size of perturbations.

To model a two component universe composed of  a relativistic plasma (RP) and
a perfect fluid (PF), we introduce the
mixing-factor
\begin{equation}
\alpha ={\rho_{\rm RP}\over \rho_{\rm RP} + \rho_{\rm PF}} \ .
\end{equation}
Gauge-invariant matter variables $X$ are then decomposed as
\begin{equation}
X = \alpha X_{\rm RP} + (1-\alpha) X_{\rm PF} \ ,
\end{equation}
whereas gauge-invariant metric perturbations $Y$ are given by
\begin{equation}
Y = Y_{\rm RP} + Y_{\rm PF} \ .
\end{equation}
This split is possible since we shall consider only small
linearized perturbations.

\subsubsection{Scalar perturbations}
The scalar or density perturbations obey the equations
\begin{mathletters}
\label{phipi}
\begin{eqnarray}
\label{em}
{x^2\over 3}\Phi &=& \epsilon_m \\
\label{pit}
x^2 \Pi &=& \pi_T^{(0)}\ ,
\end{eqnarray}
\end{mathletters}
with $\epsilon_m$ being the density contrast on hypersurfaces that
represent everywhere the
local restframe of matter, and
$\pi_T^{(0)}$ is the anisotropic pressure. $\Phi$ and $\Pi$ are the
metric potentials, related to Bardeen's definition by
\begin{eqnarray*}
\Phi_H &=& {1\over 2} \Phi \\
\Phi_A &=& - \Pi - {1\over 2} \Phi \ .
\end{eqnarray*}
For our purposes
another variable for the density contrast turns out to be useful
\begin{equation}
\label{eg}
\epsilon_g = \epsilon_m - \frac 4x v_s^{(0)} \ .
\end{equation}
The matter velocity $v_s^{(0)}$ (related to the amplitude of the
shear \cite{Bardeen}) is given
by the solution of
\begin{equation}
\label{scon}
v_s^{\prime (0)} + {1\over x} v_s^{(0)} = {1\over 4} \left(
\epsilon_m + \eta - {2\over 3} \pi_T^{(0)} \right) - {1\over 2}
\Phi - \Pi.
\end{equation}
The quantity $\eta$ is the entropy perturbation ($\eta = 0$ corresponds
to isentropic perturbations). It appears as source term in
the trace of the perturbed
Einstein equations
\begin{mathletters}
\label{scalar}
\begin{equation}
\label{trace}
x^2\left( \Phi^{\prime\prime} + {4\over x} \Phi^{\prime} +
{1\over 3}\Phi + {2\over x} \Pi^{\prime} - {2\over 3} \Pi
\right) = - \eta \ ,
\end{equation}
whereas in the 0-0 components we encounter
the density contrast $\epsilon_g$
\begin{equation}
\label{00}
(x^2 + 3)\Phi + 3x\Phi^{\prime} + 6\Pi = 3\epsilon_g \ .
\end{equation}
\end{mathletters}
Here and in the following a prime denotes the derivative with respect to $x$.

\subsubsection{Vector perturbations}

For vector (rotational) perturbations
\begin{equation}
\label{vector}
{x^2\over 8} \Psi = v_c
\end{equation}
relates the metric
(frame dragging) potential $\Psi$ to the matter velocity $v_c$ (relative
to the normal of constant-$\tau$ hypersurfaces), which is proportional to
the amplitude of the vorticity \cite{Bardeen}.
The anisotropic pressure is given through
\begin{equation}
\label{vcon}
v_c^{\prime} = - {1\over 8} \pi_T^{(1)}.
\end{equation}

\subsubsection{Tensor perturbations}

Tensor perturbations describe the propagation of gravitational waves,
their  evolution equation has the anisotropic pressure as source term
\begin{equation}
\label{tensor}
x^2 \left( H^{\prime\prime} + {2\over x} H^{\prime} + H \right)
= \pi_T^{(2)} \ .
\end{equation}

\section{Perturbative thermal field theory}

\subsection{Fluctuations from thermal $\lambda \phi^4$ theory}

To obtain the expressions for the matter variables we use the
thermal-field-theoretic
approach \cite{Kraemmer} instead of the usual kinetic approach.
The perturbation of the energy momentum tensor
\begin{eqnarray}
\delta T_{\ \nu}^{\mu}(x) &=& \int_{x^\prime} {\delta T^{\mu}_{\ \nu}(x)\over
\delta g_{\alpha\beta}(x^{\prime})} \delta g_{\alpha\beta}(x^{\prime})
\nonumber \\
\label{delT}
&=& {2\over \sqrt{-g(x)}} \int_{x^\prime}
\Pi^{\mu\ \alpha\beta}_{\ \nu}(x,x^\prime) \delta
g_{\alpha\beta}(x^{\prime}) - \left[\frac 12 T_{\ \nu}^{\mu} g^{\alpha\beta}
+ T^{\mu\alpha} \delta^{\beta}_{\nu} \right](x) \delta g_{\alpha\beta}(x)
\end{eqnarray}
is related to the gravitational polarization tensor
\begin{equation}
\Pi^{\mu\nu\alpha\beta}(x,x^\prime)\equiv
{\delta^2\Gamma\over\delta g_{\mu\nu}(x)\delta g_{\alpha\beta}(x^\prime)} =
\frac 12 {\delta \left( \sqrt{-g(x)}T^{\mu\nu}(x)\right)\over \delta
g_{\alpha\beta}(x^\prime)} \ .
\end{equation}
In the high-temperature domain the effective action is conformally
invariant \cite{NRS2,Rebhan91},
i.e. $\Gamma[S^2 g_{\mu\nu}] = \Gamma[g_{\mu\nu}]$, therefore the polarization
tensor in the conformally flat Einstein-de Sitter background reads
\begin{equation}
\left. \Pi^{\mu\nu\rho\sigma}(x,x^{\prime})
\right|_{g=S^2\eta} = S^{-2}(\tau)
\int {d^4 k \over (2\pi)^4}
e^{\imath k (x - x^{\prime})} \left.
\tilde{\Pi}^{\mu\nu\rho\sigma}(k)
\right|_{\eta} S^{-2}(\tau^{\prime}) \ .
\end{equation}
Additionally we have performed a Fourier transformation to momentum space,
where standard thermal-field-theoretic methods apply.
Due to the conformal
invariance and the invariance under general coordinate transformations,
the gravitational polarization tensor has only three independent components.
We choose
them to read
\begin{equation}
A(Q)\equiv\tilde\Pi^{0000}(Q)/\rho,\quad
B(Q)\equiv\tilde\Pi^{0\mu}{}_\mu{}^0(Q)/\rho,\quad
C(Q)\equiv\tilde\Pi^{\mu\nu}{}_{\mu\nu}(Q)/\rho.
\end{equation}
The explicit expressions for $A, B$, and $C$ have been calculated in
\cite{NRS2} to ${\cal O}(\lambda^{\frac32})  $. The contributions through
order $\lambda$
consist of ``hard thermal loops'', i.e., diagrams that are
dominated by hard loop momenta. These correspond to thermal fluctuations and
could have been calculated by kinetic theory as well (despite the
collision term that cannot be derived by first principles).
The order $\lambda^{\frac32}$ stems
from the necessary resummation \`a la Braaten-Pisarski \cite{BP}
where propagators
and vertices of the scalar particles have to be dressed in order to avoid
infra-red singularities. The resummation involves a particular (infinite)
subclass of
Feynman diagrams of ordinary perturbation theory which presumably lie far
beyond the scope of kinetic theory. $A,B,C$ and their Fourier transforms
are listed in Appendix A.

\subsubsection{Scalar perturbations}
{From} (\ref{delT}) and the definition of the gauge-invariant matter
variables \cite{Bardeen}
it follows that
\begin{mathletters}
\label{mv}
\begin{eqnarray}
\eta_{\rm RP} &=& 0 \;,\\
\epsilon_{g\ \rm RP} &=& - 2\Phi + 4 {\cal F}[A - \frac14]*(\Phi + \Pi) \;,\\
v_{s\ \rm RP}^{(0)} &=&
3\imath {\cal F}[\omega (A - \frac14)]*(\Phi + \Pi)\;, \\
\pi_{T\ \rm RP}^{(0)} &=& - 18 {\cal F}[(\omega^2 - \frac13)
(A - \frac14) + \frac13]*(\Phi + \Pi) \;.
\end{eqnarray}
\end{mathletters}
The $*$ denotes the convolution $(g * f) (x) = \int^x dx^\prime
g(x-x^\prime) f(x^\prime)$ and the operator ${\cal F}$ defines
the Fourier transformation
\begin{equation}
{\cal F}[g](x) = \lim_{\gamma \to 0^+} {1\over 2\pi}
\int_{-\infty +\imath \gamma}^{\infty + \imath \gamma} d \omega
e^{-\imath \omega x } g(\omega) \ .
\end{equation}
The particular choice of the integration contour corresponds to
retarded boundary conditions.

After performing the Fourier transformations (Appendix A) the
matter variables are written in terms of the integral kernel
\begin{eqnarray}
K^{(0)}(x) = &\biggl[& j_0(x)
+ {5 \lambda \over 8 \pi^2} \left(2\kappa^\prime - j_0 -\cos\right)(x)
\nonumber \\
&& + {15 \over 8} \left(\lambda\over \pi^2\right)^{\frac32}
\left(\cos + \frac x3 J_{-1} + 2 J_0 + \frac 1x J_1 + \frac 43 j_2
- \frac 23 j_0 \right. \nonumber \\
\label{K0}
&&\qquad \left. - 2 \nu^{\prime\prime} - 4 \nu - \kappa^\prime
- \kappa^{\prime\prime\prime} - \frac 16 \xi \right)(x) \biggr] \ ,
\end{eqnarray}
\begin{mathletters}
according to
\begin{eqnarray}
\epsilon_{g\ \rm RP} &=& 2\Phi+4\Pi-4 K^{(0)}*(\Phi + \Pi)^\prime \\
\label{vsk}
v_{s\ \rm RP}^{(0)} &=& 3  K^{(0)\,\prime}*(\Phi + \Pi)^\prime \\
\label{pitk}
\pi_{T\ \rm RP}^{(0)} &=&
- 6  (K^{(0)} + 3 K^{(0)\,\prime\prime})*(\Phi + \Pi)^\prime
\end{eqnarray}
\end{mathletters}
At the origin this kernel behaves like
\begin{equation}
K^{(0)}(0^+) = 1, \qquad K^{(0)\prime}(0^+) = 0 \ .
\end{equation}
The scalar matter variables (\ref{mv}) couple only via
the sum of the metric potentials $\Phi + \Pi \equiv \Phi_N$
to the nontrivial component $A$ of the polarization tensor.
$\Phi_N$ is the only scalar contribution to the electric part of the
Weyl curvature tensor \cite{Ellis} and can be interpreted as the generalization
of the Newton gravity potential. There is no scalar contribution in the
magnetic part of the Weyl tensor. Were it not for the conformal symmetry of
the effective action, the additional components of the polarization
tensor would also
couple to $\Pi$, the potential for the anisotropic pressure.

For vanishing entropy perturbations, Eq. (\ref{trace})  can be
integrated for $\Phi$,
\begin{eqnarray}
\Phi (x) &=& - \frac{2}{3 x} \int^x dx^{\prime} \left(x^\prime
\cos ( x-x^{\prime}  )  +
4 \sin (x - x^{\prime} ) \right) \Phi_N^{\prime} (x^{\prime} ) \nonumber \\
\label{intPhi}
& & + \left. \frac{2}{3x}
\left(x^\prime \cos (x-x^{\prime}) + \sin (x - x^{\prime}) \right)
\Phi_N (x^{\prime} )\right|^x
+ C_1 \frac{\sin (x)}{x} + C_2 \frac{\cos (x)}{x}
\end{eqnarray}
where $C_1,C_2$ are integration constants. Together with
Eqs.\ (\ref{pit}) and (\ref{pitk}) we can also obtain a single equation
determining
$\Phi_N$ (see Eq.~(\ref{PhiN})).

\subsubsection{Vector perturbations}
In a similar way
\begin{equation} \label{vect}
v_{c\ \rm RP} = -\Psi + \frac32 {\cal F}[(\omega^2 -1)(A - \frac14) - B]*\Psi
\end{equation}
is derived. With
\begin{equation}
\label{K1}
K^{(1)}(x) = \biggl[ - \frac13
(1+\frac{15 \lambda}{8 \pi^2 } - \frac{15 \lambda^{3/2} }{2 \pi^3 })
(j_0 + j_2)(x)
+ {5 \lambda \over 8 \pi^2} j_0(x)
- {15\over 8}\left(\lambda\over \pi^2\right)^{\frac 32}
\left( \frac13 J_0 + j_0 \right) (x)
\biggr]
\end{equation}
and
\begin{equation}
K^{(1)}(0^+) = - \frac13 \ , \qquad K^{(1)\prime}(0^+) = 0 \ ,
\end{equation}
the velocity amplitude is given by
\begin{equation}
\label{vk}
v_{c\ \rm RP} =  - 3 ( K^{(1)\,\prime}) * \Psi \ .
\end{equation}

\subsubsection{Tensor perturbations}

The evaluation of (\ref{delT}) for the anisotropic pressure
leads to
\begin{eqnarray} \label{tk}
\pi_{T\ \rm RP}^{(2)}
&=& 3 {\cal F}[(\omega^2 - 1)^2 (A - \frac14)
- 4 (\omega^2 -1) B + 2 C - {11\omega^2\over 3} + 3] * H \nonumber\\
&=& 3 ( K^{(2)}) * H^\prime
\end{eqnarray}
with the kernel
\begin{equation}
\label{K2}
K^{(2)}(x) = \biggl[ - 8 {j_2(x)\over x^2}
 (1+\frac{15 \lambda}{8 \pi^2 } -
\frac{15 \lambda^{3/2} }{2 \pi^3 } )
+ {5 \lambda \over \pi^2}{j_1(x)\over x}
- 5 \left(\lambda\over \pi^2\right)^{\frac 32}
 \left(\frac{J_1}{x} +  j_0 + j_2\right)(x) \biggr] .
\end{equation}

\subsection{Initial conditions}

To specify the dynamical equations for cosmological perturbations
completely, initial conditions for the metric potentials are needed. We
fix them in the limit $x_0\to 0$,
because this will allow us to derive exact solutions of the
integro-differential equations in terms of generalized power series
\cite{Rebhan94}. The initial conditions are given by a set of numbers
$\gamma^{(a)}_n$. These are related to the n-th momenta of the
particle's distribution function \cite{Rebhan94}. Here they show up in the
convolution integrals because the lower integration boundary $x_0$
truncates the support of the metric potentials to a half space.
This is done in the convolution integrals by the replacement
\begin{equation}
\label{ic}
Y^\prime(x) \to Y^\prime(x)\theta(x-x_0) +
\sum_{n=0}^{\infty} \gamma^{(a)}_n \delta^{(n)}(x-x_0) \ ,
\end{equation}
where $Y$ stands for $\Phi_N , \Psi$, and $H$, respectively.

As is well known from simple forms of matter like perfect fluids
\cite{Lifshitz} and from work by Zakharov and Vishniac \cite{Zakharov}
on collisionless matter,
there are two branches of solutions, regular and singular ones.
This is also the case for self-interacting
plasmas. We will concentrate on regular solutions
in our evaluations,
but will nevertheless sketch the behaviour of the singular solutions
below. The singular solutions necessarily
violate the assumption of a Friedmannian
singularity, but they are relevant when fitting the evolution of
cosmological perturbations
to a previous epoch at some small nonvanishing value of $\tau$
\cite{Grishchuk,Deruelle}. A detailed discussion of singular solutions for
collisionless matter is given in \cite{Rebhan94}.
Due to geometrical effects \cite{Zakharov},
the singular solutions permit superhorizon
oscillations. This effect depends on the ratio $\alpha$ only, and is not
at all sensitive to the $\gamma_n^{(a)}$.
Only the normalization of the
singular part of the solutions has to be fixed by an initial condition.

The regular solutions are determined by the $\gamma_n^{(a)}$. We will
restrict our attention to isentropic (adiabatic) perturbations, which
may be left over from an earlier inflationary epoch.
We will neglect all $\gamma_n^{(a)}$ with $n>2$. This is motivated by the
findings of \cite{Rebhan94} that these quantities are related to the
higher moments of the kinetic distribution function. The restriction
to $n\le2$ means that we fix the momenta which directly
occur in the energy-momentum
tensor and set all others to zero initially.

\subsubsection{Scalar perturbations}

{From} (\ref{phipi}) and (\ref{scon}) the small $x$ behaviour of the matter
variables
\begin{eqnarray}
\epsilon_m &\sim& x^2 \Phi(0)\nonumber \\
v_s^{(0)} &\sim& x (\Phi(0)+2\Pi(0))\nonumber \\
\pi_T^{(0)} &\sim& x^2 \Pi(0)\nonumber
\end{eqnarray}
follows. Therefore, $v_s^{(0)}(0^+)$ and $\pi_T^{(0)}(0^+) = 0$
for regular solutions.
This means that
the r.h.s. of (\ref{vsk}) and (\ref{pitk}) after performing (\ref{ic})
have to vanish as well, i.e.,
\begin{eqnarray}
\sum_{n=0}^\infty \gamma_n^{(0)}
K^{(0)(n+1)}(0^+) &=& 0 \\
\sum_{n=0}^\infty \gamma_n^{(0)}
(K^{(0)} + 3 K^{(0)\prime\prime})^{(n)}(0^+) &=& 0  \ .
\end{eqnarray}
Due to
\begin{equation}
\label{intc}
K^{(0) \prime}(0^+) = (K^{(0)} + 3 K^{(0)\prime\prime})(0^+) = 0
\end{equation}
we are free to choose any $\gamma_0^{(0)}$. It is remarkable that these
relations hold for collisionless matter and are not changed by the addition
of weak self-interactions.
A completely arbitrary choice of the $\gamma_n^{(0)}$ is not
possible. In the following we choose $\gamma_0^{(0)} \neq 0$ and all higher
$\gamma_n^{(0)}$ vanishing. This corresponds to the usual choice for
isentropic initial conditions, e.g. \cite{Schaefer}.

In the above mentioned equation for $\Phi_N$ the arbitrary constants
$C_1$ and $C_2$ remain to be fixed.
Initial conditions are included by the replacement (\ref{ic})
which yields, together with Eqs. (\ref{intPhi},\ref{pit},\ref{pitk}),
the integral equation for $\Phi_N$
\begin{eqnarray}
\label{PhiN}
\Phi_N(x)=
&&- \frac{2}{x} \int_0^{x} dx^{\prime} \left( x^\prime
 \cos (x-x^{\prime}  )  +
4 \sin (x - x^{\prime} ) \right) \Phi_N^{\prime} (x^{\prime} ) \nonumber \\
\label{phiN}
&&- \frac{18 \alpha}{x^2} \int_0^x  d x^\prime
 ( K^{(0)} + 3  K^{(0)\,\prime\prime})(x-x^\prime)
\Phi_N^\prime (x^\prime) + \mbox{i.c.}
\end{eqnarray}
with the initial conditions
\begin{eqnarray}
\mbox{i.c.} = &&
\frac{ \sin x}{ x} \left(3 C_1 - 2 \Phi_N(0) + 4
\sum_{n=0}^\infty \gamma_{2n} (-)^n (n-2)  \right) +
\frac{ \cos x}{x} \left(3  C_2 +2
\sum_{n=0}^\infty \gamma_{2n+1} (-)^n (2 n-3)  \right) \nonumber \\
&&- \frac{18 \alpha}{x^2} \sum_{n=0}^\infty \gamma_n
( K^{(0)} + 3  K^{(0)\,\prime\prime} )(x) )^{(n)}.
\end{eqnarray}begin
For regular solutions, $C_2$ is determined by requiring that the
term in the second round brackets vanishes, and $C_1$ is related
to the initial value $\Phi_N(0)$ by the
constant term in the series in $x$.

The metric perturbation $\Phi_{\rm RP}$ is given through
\begin{equation}
\label{phiRP}
\Phi_{\rm RP} = {3 \alpha \over x^2} \left( \epsilon_{g\ \rm RP} + \frac 4x
v_{s\ \rm RP}^{(0)}\right) \ .
\end{equation}
The regularity requirement establishes a relation between $\Phi_N(0), \Pi(0)$
and the $\gamma_{2n}^{(0)}$ coefficients. Using this relation together with
the expressions for $\Phi_N(0)$ and $\Pi(0)$, $C_1$ may be expressed as a sum
of coefficients $\gamma_{2n}^{(0)}$ (see Appendix B).

\subsubsection{Vector perturbations}

{From} (\ref{vcon}) a constant $v_{c \rm\ PF}$ follows, which is a
consequence of the Kelvin-Helmholz theorem. For a perfect fluid
alone, this forbids regular solutions at all, because it entails
a singular frame dragging potential $\Psi$. This
situation changes when collisionless or weakly
self-interacting matter
is added (a discussion of potential cosmological
consequences like primordial magnetic fields can be found in \cite{Rebhan92b}).
One can have a regular vector perturbation sustained by the
relativistic plasma alone, or
one can compensate a primordial vorticity in the perfect fluid
component, which has constant $v_{c \rm\ PF}$,
by a nonvanishing vorticity of opposite sign in the relativistic plasma, and
a growing net vorticity is generated by the nontrivial evolution of the
latter.

Eq.~(\ref{vector}) implies for small $x$
\begin{equation}
v_c = \alpha v_{c\rm\ RP} + (1-\alpha)v_{c \rm\ PF} \sim x^2 \ .
\end{equation}
Therefore, (\ref{vk}) leads with (\ref{ic}) to
\begin{equation}
\label{vPF}
v_{c \rm\ PF} = \frac{ 3 \alpha}{1-\alpha}
   \sum_{n=0}^\infty
\gamma_n^{(1)} K^{(1)(n)}(0^+) \ .
\end{equation}
As can be seen from the above formula, in the absence of a primordial
perfect fluid vorticity one has to have nonvanishing coefficients
$\gamma_n^{(1)}$, $n\ge1$ for nontrivial regular solutions.
In the presence of a perfect fluid component, we shall restrict
ourselves to nonzero
$\gamma_0^{(1)}$ and vanishing higher coefficients.

\subsubsection{Tensor perturbations}

Eqs. (\ref{tensor}) and (\ref{tk}) with (\ref{ic}) lead to
\begin{equation}
\sum_{n=0}^\infty \gamma_n^{(2)} K^{(2)(n)}(0^+) = 0 \ .
\end{equation}
Because tensor perturbations correspond to gravitational waves,
there are nontrivial
solutions for all $\gamma_n^{(2)}$
vanishing. Then the initial conditions are
specified by the amplitude $H(0)$ and its derivative $H^\prime(0)$.
This will be our choice in what follows.

\subsection{Solutions through $O(\lambda^{3/2})$}

The singular solutions to the equations for the dynamics
of cosmological perturbations are obtained with the Ansatz
\begin{equation}
Y^{(a)}(x) = x^\sigma \bar{Y}^{(a)}(x)
\end{equation}
with $\bar{Y}^{(a)}(0^+) \neq 0$ and finite ($a=0,1,2$).
This yields
\begin{equation}
\sigma = -\frac52 + a \pm
\frac12 \sqrt{1 - {\alpha\over \alpha_{\rm crit.}(\lambda)}} \ .
\end{equation}
Therefore, for $\alpha$ greater than
\begin{equation}\label{acrit32}
\alpha_{\rm crit.} (\lambda) = \frac5{32} \left(
1 - \frac54 {\lambda\over \pi^2} + {105\over 16}
 \left(\lambda\over \pi^2\right)^{\frac32}
\right)^{-1}
\end{equation}
$\sigma$ takes complex values, and thus gives rise to superhorizon
oscillations $\sim\cos([{\mathrm Im}\sigma]\ln x)$.
For small
$\lambda \geq 0$ the value of
$\alpha_{\rm crit.}$ is increased. This is as one may
expect, since in the collision-dominated case of a perfect fluid
superhorizon oscillations do not occur.
For large values of $\lambda$,
$\lambda \ge 8\pi^2/63\approx 1.25$
this trend is reversed, but there the perturbative series can no
longer be trusted, because the $\lambda^{3/2}$ correction begins
to dominate over the $\lambda^1$-term.

Turning now to the regular solutions, the power series
Ansatz detailled in Appendix B is made. This leads to recursion
relations for its coefficients that can be solved
as also given in the Appendix. These have been
evaluated with a {\em Mathematica} code \cite{Wolfram}.
On superhorizon scales
the effects of the self-interactions on the regular solutions are small.
E.g., for scalar perturbations with $\alpha = 1$ and the
above initial conditions,
\begin{equation}\label{pioeps}
{\pi_T^{(0)}\over \epsilon_m^{(0)}}(0) = - \frac 37 \left(
1 - {25\over 28} {\lambda\over \pi^2} + {75\over 16}
\left(\lambda\over \pi^2\right)^{\frac32} \mp O(\lambda^2)\right)
\end{equation}
the amplitude of the anisotropic pressure becomes smaller when the
plasma is collisional for values of $\lambda$ smaller $ (4\pi/21)^2
\sim 0.36$. Since
for tightly coupled plasmas we expect the behaviour of a
perfect fluid, i.e. the above ratio should be zero, we conclude that
the perturbative result through order $\lambda^{3/2}$ can be trusted only
for $\lambda < 0.36$.

\begin{figure}                               
\centerline{ \epsfbox{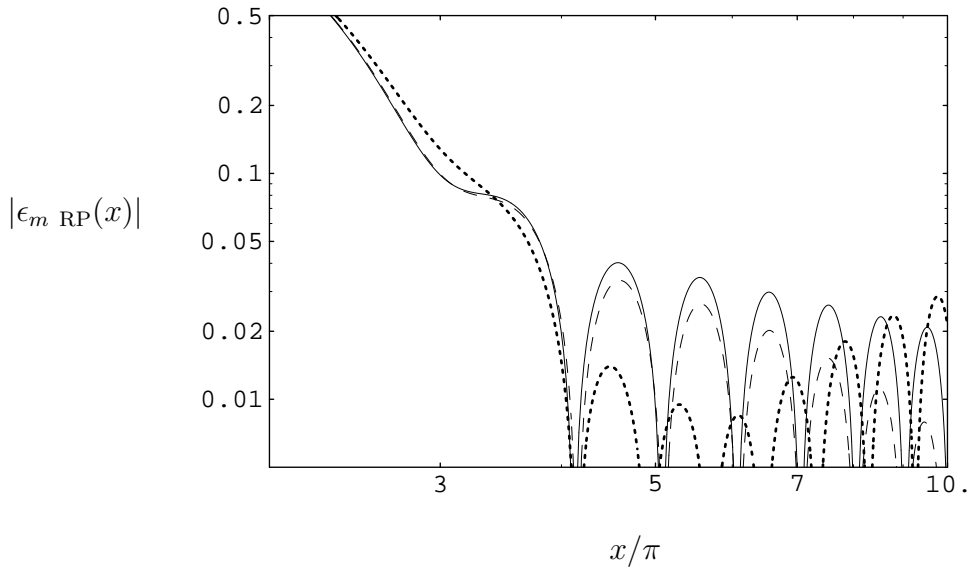} }
\vspace{-4cm}
\unitlength1cm \begin{picture}(16,3)
\put(0,4.5){$|\epsilon_{m\ {\rm RP}}(x)|$}
\put(8,0){$x/\pi$}
\end{picture} \\[30pt]
\caption{\label{f1}
The full line shows the density contrast $|\epsilon_{m\ {\rm RP}}|$
for collisionless matter in the subhorizon region. The dashed and
dotted solutions show the effect of collisional matter through
$O(\lambda^{3/2})$ with
$\lambda = 1/2$ and $1$ respectively.
}
\end{figure}

Some solutions for the density
contrast in the subhorizon region
are shown in Fig.~\ref{f1}. The three lines show
the behaviour of $|\epsilon_{m\ {\rm RP}}(x)|$
for an ultrarelativistic plasma alone. We plotted solutions
for $\lambda = 0,1/2$, and $1$. The collisionless solution (full
line) decays due to directional dispersion \cite{Boerner}.
The phase
velocity of the oscillations is the speed of light.
For very small values of $\lambda$ no visible effect occurs in the
plotted region, whereas for bigger values a considerable change in the
damping behaviour and in the phase velocity is observed. For $\lambda =1$
(dotted line) the density contrast starts to grow again at $x/\pi \sim 6$,
which in fact is associated with rather unlikely beats. Also for the
smaller value of $\lambda=1/2$ such a behaviour arises for large enough $x$.
This will be analysed later in the next section.

\section{Pad\'e improvement}

As we have seen, in order to have sizable effects from the
self-interactions of the thermal matter, we have to adopt
sizable values of $\lambda$. However, the perturbative results
quickly become unreliable with increasing $\lambda$. In order
to get a somewhat more quantitative idea of the problem and
of ways to improve this situation, we first inspect a
solvable case.

\subsection{The thermal mass in the limit $N\to\infty$}

It is well-known\cite{DolanJackiw} that in the limit $N\to\infty$
the model of Eq.~(\ref{L}) becomes exactly solvable. In this limit,
the thermal mass of the scalars is exactly given by the resummed
one-loop gap equation
\begin{equation} \label{mgap}
m^2/T^2={6\lambda\over\pi^2}\int_{m/T}^\infty dx \, { \sqrt{x^2-m^2/T^2}
\over e^x-1} .
\end{equation}

The first two terms of the perturbative series
\begin{equation} \label{mpert}
m^2/T^2=\lambda-\frac3\pi \lambda^{3/2}+O(\lambda^2\ln\lambda)
\end{equation}
are naturally a good approximation for very small $\lambda$,
but apparently breaks down when $\lambda\approx1$. Indeed,
for $\lambda=1$, Eq.~(\ref{mpert}) gives $m^2/T^2\approx 0.05$, which
is an order of magnitude too small when compared with the
result following from Eq.~(\ref{mgap}), which yields $m^2/T^2\approx  0.53$.
However, rewriting  Eq.~(\ref{mpert}) in the perturbatively equivalent
way of
\begin{equation} \label{mpade}
m^2/T^2=  {\lambda\over1+ \frac3\pi \lambda^{1/2} }+O(\lambda^2\ln\lambda)
\end{equation}
considerably extends the range of $\lambda$ over which the
first two terms of the perturbative series give a faithful
picture of the actual behaviour. For $\lambda=1$,  Eq.~(\ref{mpade})
yields  $m^2/T^2\approx  0.51$, which is only a few percent too small.

\begin{figure}                               
\centerline{ \epsfxsize=4in \epsfbox[68 240 540 560]{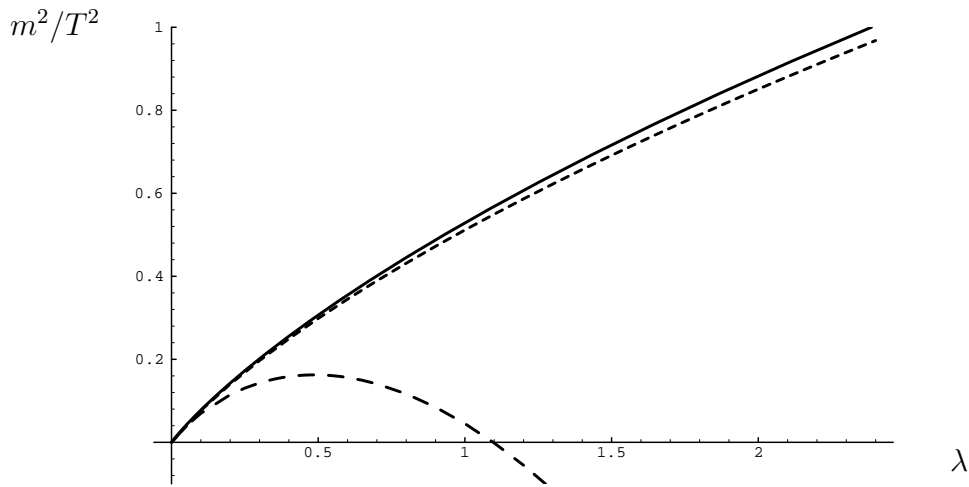} }
\vspace{-4cm}
\unitlength1cm \begin{picture}(16,3)
\put(1,5.3){$m^2/T^2$}
\put(13.5,-0.5){$\lambda$}
\end{picture} \\[30pt]
\caption{\label{lam2}
$m^2/T^2$ as a function of $\lambda$ in the exactly solvable case
of $N\to\infty$ (full line). The long-dashed curve is the
perturbative result (4.2); the short-dashed one the
Pad\'e-improved (4.3).
}
\end{figure}                                               

Of course, for still larger values of $\lambda$, also Eq.~(\ref{mpade})
becomes increasingly imprecise, but the improvement over Eq.~(\ref{mpert})
remains striking.
What we have done by going  from  Eq.~(\ref{mpert}) to Eq.~(\ref{mpade})
can be viewed as replacing the first terms of a power series in
$\sqrt{\lambda}$
\begin{equation}
X=a+b \lambda^{1/2}+c\lambda+d\lambda^{3/2}+\ldots
\end{equation}
by its (2,1)-Pad\'e approximant (again in powers of
$\sqrt{\lambda}$)\cite{Pade}
\begin{equation}
X={\alpha+\beta\lambda^{1/2}+\gamma\lambda \over 1+\delta\lambda^{1/2}}
+\ldots
\end{equation}
In the case of the thermal mass, $a=0$ since we have no tree-level
mass to start with, and $b=0$ since the plasmon effect is down by
one-half order with respect to the  perturbative one-loop one.
Extending this procedure to other quantities, we thus still have $b=0$,
so that the results through order $\lambda^{3/2}$
determine the four parameters of the corresponding (2,1)-Pad\'e approximants,
\begin{equation}  \label{abcd}
\alpha=a,\quad
\beta=-ad/c,\quad
\gamma=c,\quad
\delta=-d/c .
\end{equation}

\subsection{The resummed 1-loop kernel}

In our applications, the accuracy of the
perturbative result will not only deteriorate
when $\lambda$ is increased, but also when $x$ becomes too large,
as we have seen in Fig.~\ref{f1}.
This is so because the asymptotic behaviour of the Fourier transforms
(\ref{K0},\ref{K1},\ref{K2}) with $x\to\infty$ is such
that for any small but finite
value of $\lambda$ the correction of order $\lambda^{3/2}$ in $K^{(a)}$
eventually
overtakes the $\lambda^1$ term
which in turn overtakes the lowest order term. Apparently,
this signals a breakdown of perturbation theory at large $x$.

The origin of the difference in the asymptotic behaviour in $x$
comes from an increasingly singular behaviour of the discontinuity
of the functions $A$, $B$, and $C$ at higher orders in $\lambda$.
For example, from (\ref{Ar}) one notices that at lowest order the
discontinuity of $A$ across the branch cut between $\omega=\pm1$
is a constant; at order $\lambda^1$ it is logarithmically singular
at $\omega=\pm1$ and in addition, there are now simple poles at
these points; at order $\lambda^{3/2}$ the singularities are
still worse. Since these singularities occur at the end points
of the integration region contributing to the Fourier transform,
they become dominant for the large $x$ behaviour of the latter.

On the other hand, these singularities at the light-cone
should not exist at all since the originally massless scalars
have acquired thermal masses. Indeed, keeping the thermal masses
in the integrals without expanding them on account of them
being proportional to $\lambda$, shows that the complete
discontinuity is smooth at $\omega=\pm1$ and that also the
simple poles there are spurious \cite{NRS1}.
For instance, the one-loop contribution to $K^{(0)}(x)=j_0(x)$
is modified by including the thermal mass in the scalar propagators
according to
\begin{eqnarray}\label{R18}
K_1^{\rm res.}(x)= - && \int_{-1}^1 d\omega e^{-i\omega x}\\ \nonumber
\times
\int_{m/\sqrt{1-\omega^2}}^\infty && dp\,p^4
{d\over dp}\left({  1\over \exp (p/T) -1 } \right)
\left/ \left( {8\pi^4 T^4\over15} \right)  \right.
\end{eqnarray}
up to terms whose amplitude is suppressed by explicit powers of $\lambda$.
With a non-zero $m=\sqrt{\lambda}T$,
the integrand is now seen to vanish at $\omega=\pm1$. Instead of
being a constant, it vanishes at the endpoints of the branch cut
together with all its derivatives, but rapidly recovers the bare one-loop
value away from the light-cone. This is a negligible
effect for small $x$,
but the large $x$ behavior is changed completely.

Eq.\ (\ref{R18}) can be evaluated by a Mellin transform\cite{Dav}
which yields
\begin{eqnarray}\label{R19}
&&K_1^{\rm res.}(x)=
{\sqrt\pi\over2}{15\over 8 \pi^4} \times\\ \nonumber
&&\sum_{k=0}^\infty \lambda^{k/2}
{(-1)^{1+k}(k-4)\zeta(k-3)\over (2\pi)^{k-4} \Gamma(1+k/2)}
\left(2\over x\right)^{{(  1-k)/ 2}} J_{{ (1-k )/ 2}}(x) ,
\end{eqnarray}
where in the term with $k=4$ one has to substitute $(k-4)\zeta(k-3)\to1$.
For small $x$,
\begin{equation}\label{K1rexp1}
K_1^{\rm res.}(x)
= \left( j_0(x)-{5\lambda\over 8\pi^2}\cos(x) \right)
+O(\lambda^{3/2})
\end{equation}
is a good approximation;
for $x\gg1$, on the other hand, the complete function
$K_1^{\rm res.}(x)$ turns out to
decay even faster than $j_0(x)$, oscillating with a reduced
phase velocity
\begin{equation}
v=1-{5\lambda\over 8\pi^2}+O(\lambda^{3/2}).
\end{equation}

A better approximation is thus obtained by modifying
$$
K^{(0)}_{1}(x) \propto j_0(x) \to j_0((1-\frac{5\lambda}{8\pi^2})x).
$$
This indeed resums and thus removes the terms of order $\lambda$
in $K^{(0)}$ which are the most
dominant for large $x$, leaving only such which are logarithmically
larger than $K^{(0)}_1$. However, the kernel $K^{(0)}$ has to
satisfy (\ref{intc})
in order that the equation for scalar cosmological perturbations
remains integrable. In Ref.~\cite{NRS1} we have found that this
can be fulfilled by adjusting the order $\lambda$ term in $K^{(0)}$ so
that it contains the same
phase velocity as the modified $K^{(0)}_1$ and correcting the prefactors
without modifying the lowest orders in $\lambda$.

However, it is difficult to see how the effect of the higher terms
in Eq.~(\ref{R19}) can be accounted for by similarly simple
modifications. The functions that come with higher
powers of $\lambda$ are in fact more and more {\em diverging}
with $x$ and only the infinite sum is a decaying
and regularly oscillating function ---
any truncation results in a function that oscillates with
interferences before exploding eventually.

Unfortunately, we are able to compute only the first few terms
of the contributions to the kernel, Eq.~(\ref{R18}) being only one
particularly simple contribution. In order to improve the
behaviour of our results at large values of $x$, we propose
to use the same Pad\'e approximation that was working
rather well above. The most obvious way for doing that
would be to turn the coefficients in Eq.~(\ref{abcd}) into
functions of $x$. However, this turns out to violate the
integrability constraints on $K^{(a)}$. A method which manifestly
respects the latter is to perform a Pad\'e improvement on
each of the Taylor coefficients of $K^{(a)}$ as a function of $x$.

\begin{figure}                               
\centerline{ \epsfxsize=5in \epsfbox[68 230 540 560]{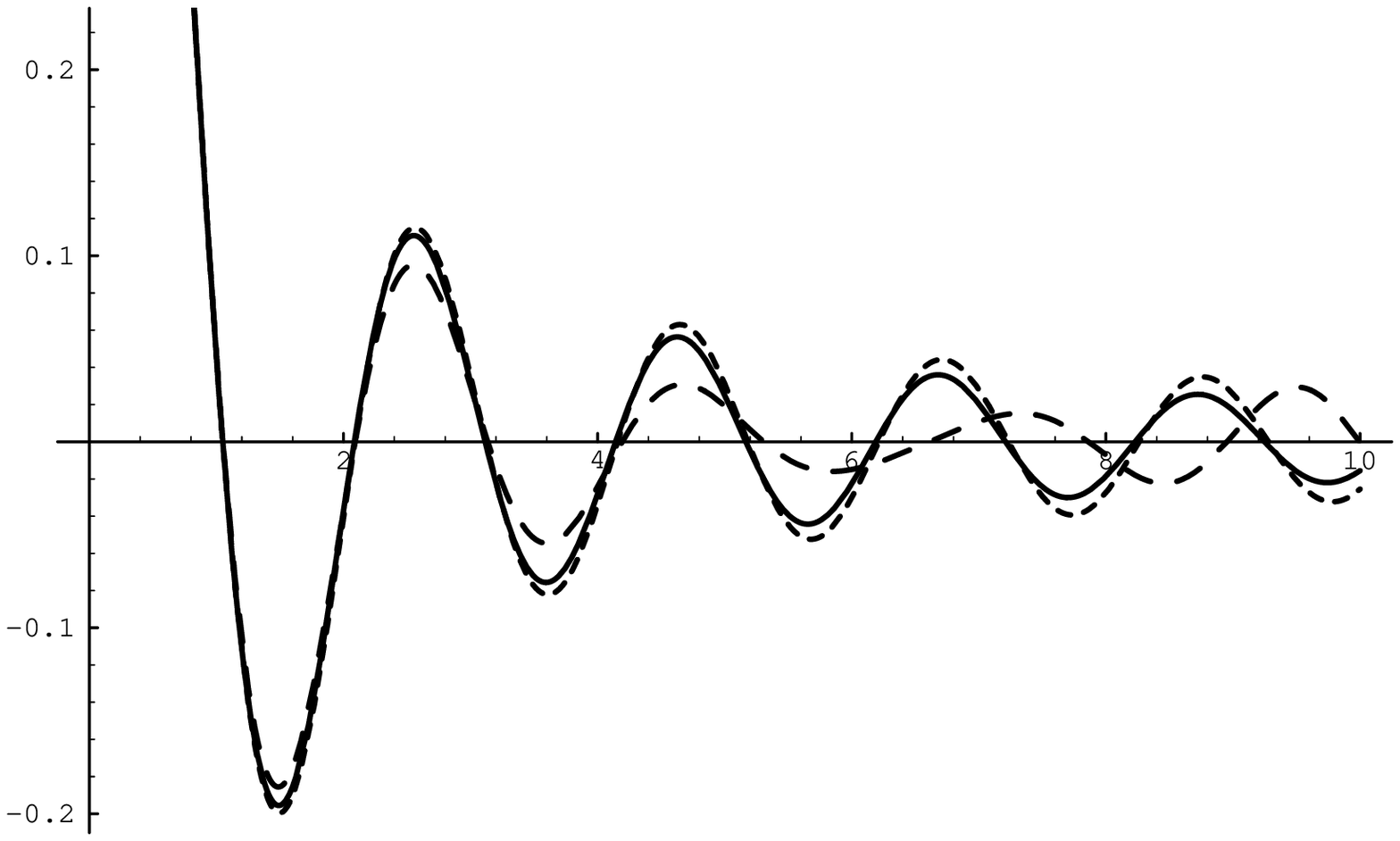} }
\vspace{-6cm}
\unitlength1cm \begin{picture}(16,5)
\put(0,7.2){$K_1^{\rm res.}(x)$}
\put(14.5,3){$x/\pi$}
\end{picture} \\[30pt]
\caption{\label{padek}
The function $K_1^{\rm res.}(x)$ (full line) for $\lambda=1$
and two perturbative approximations: The long-dashed line gives
the first three terms of its perturbative expansion;
the short-dashed one its Pad{\'e} improved version.
}
\end{figure}

In order to test this procedure, we have applied it to the first three
terms of the expanded kernel (\ref{R19}). With $\lambda=1$, the
result is shown in Fig.~\ref{padek}. There the full line gives the
complete function (\ref{R18}) and the long-dashed line shows the
truncated series through $\lambda^{3/2}$. The latter deviates
quickly from the complete function and after few oscillations
is completely off. The Pad\'e improved function, where
each term in the power series in $x$ is replaced by
a (2,1)-Pad\'e approximant, behaves instead much more
regularly and is quite close to the complete function for
all values of $x$.

We therefore expect that the analogous procedure for
improving the kernels we have obtained perturbatively up to and including
order $\lambda^{3/2}$ allows us to extend the range in both,
$\lambda$ and $x$, where the solutions to the corresponding
equations for cosmological perturbations can be trusted,
to about $\lambda\lesssim1$ and $x/\pi\lesssim10$.

\section{Pad\'e-improved solutions}

On superhorizon scales the perturbative results became obviously unreliable
already at moderate values of $\lambda$ because, among others, the
$O(\lambda^{3/2})$-contributions in
Eq.~(\ref{acrit32}) stopped $\alpha_{\rm crit.}$ from
increasing beyond $\lambda\approx1.25$. With the Pad\'e-improved results
we have
\begin{equation}
\alpha_{\rm crit.}(\lambda) = \frac5{32}{4 + 21{\lambda^{\frac12}\over\pi}
\over 4 + 21 {\lambda^{\frac12}\over\pi} - 5 {\lambda\over \pi^2}},
\end{equation}
which shows an ever-increasing behaviour up to very large $\lambda$.
Likewise the Pad\'e-improved version of Eq.~(\ref{pioeps}) is a
monotoneous function. In both cases, the results for the weakly-interacting
plasma remains far from the perfect-fluid ones for $\lambda\sim1$; only
for $\lambda\gtrsim 10^2$ these would be reached, where a perturbative
treatment is certainly inadequate.

In the following we shall inspect our Pad\'e-improved solutions for $\lambda=1$
and $x/\pi\ge10$.

\subsection{Scalar perturbations}

\begin{figure}                               
\centerline{ \epsfbox{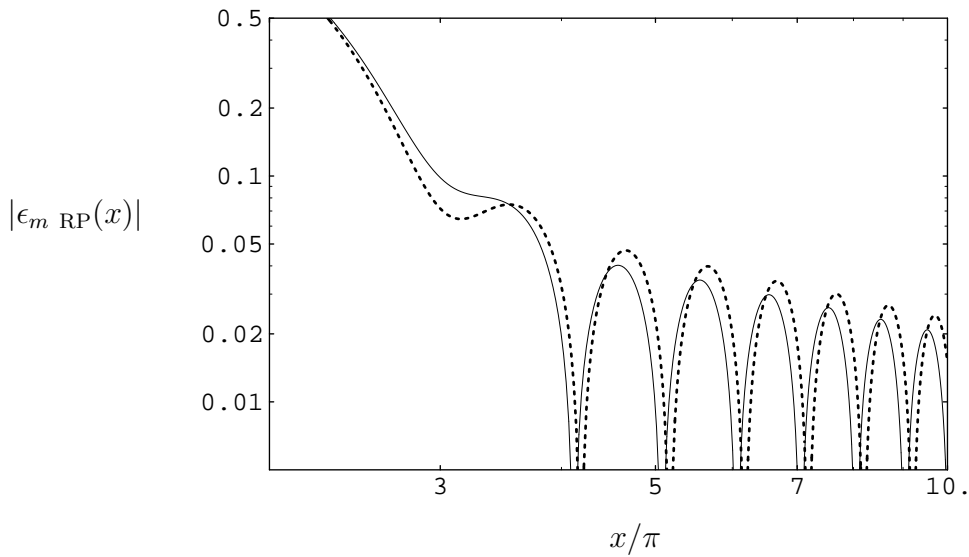} }
\vspace{-4cm}
\unitlength1cm \begin{picture}(16,3)
\put(0,4.5){$|\epsilon_{m\ {\rm RP}}(x)|$}
\put(8,0.2){$x/\pi$}
\end{picture} \\[5pt]
\caption{\label{f4}
The density contrast is shown for collisionless matter (full line)
and for an ultrarelativistic
plasma with $\lambda = 1$ (dotted line), which is
the Pad\'e-improved solution.
}
\end{figure}

In Fig.~\ref{f4}, the density contrast associated with a scalar
perturbation in a pure relativistic plasma is given for the
interacting and the collisionless case. The difference turns out
to be moderate so that a perturbative treatment seems justified.
The main effect turns out to be a somewhat decreased phase velocity
and a somewhat diminished exponent in the power-law decay.
This is exactly what one would expect in view of the behaviour
of density perturbations in the presence of a perfect fluid,
where the phase velocity equals $1/\sqrt3$ and where damping
through directional dispersion is inoperative.

\begin{figure}                               
\centerline{ \epsfbox{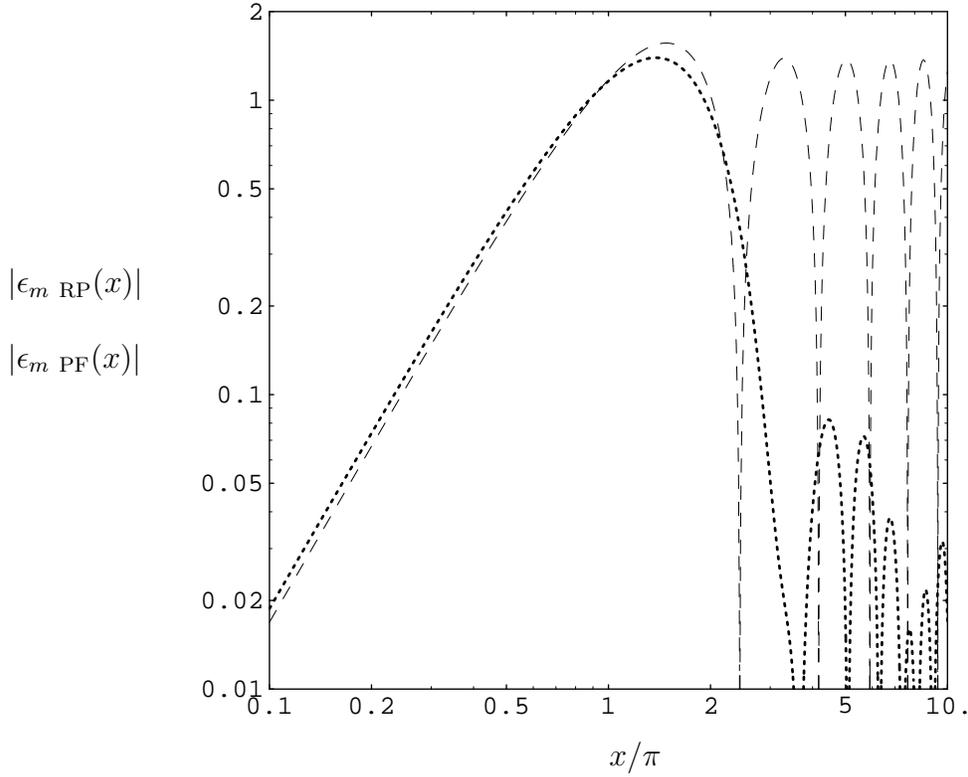} }
\vspace{-4cm}
\unitlength1cm \begin{picture}(16,3)
\put(0,5){$|\epsilon_{m\ {\rm RP}}(x)|$}
\put(0,4){$|\epsilon_{m\ {\rm PF}}(x)|$}
\put(8,-1.3){$x/\pi$}
\end{picture} \\[40pt]
\caption{\label{f5}
Density perturbations for
a mixture of a perfect fluid (dashed line)
and an ultrarelativistic plasma (dotted line)
with $\alpha = 1/2$ and $\lambda = 1$.
}
\end{figure}

\begin{figure}                               
\vspace{-1cm}
\centerline{\epsfbox{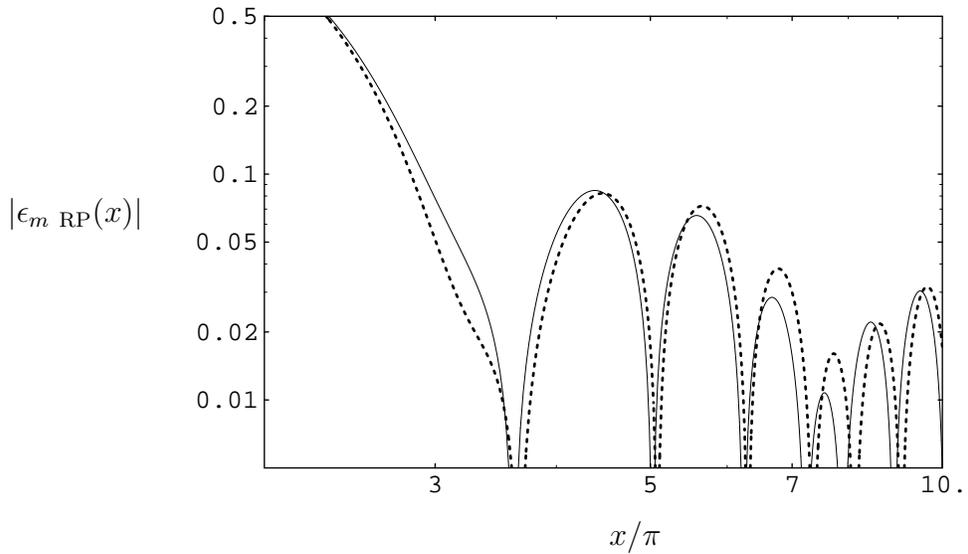}}
\nopagebreak
\vspace{-4cm}
\nopagebreak
\unitlength1cm \begin{picture}(16,3)
\put(0,4.5){$|\epsilon_{m\ {\rm RP}}(x)|$}
\put(8,0.2){$x/\pi$}
\end{picture} \\[5pt]
\nopagebreak
\caption{\label{f6}
For the same mixture as in Fig.~5 the subhorizon perturbations
of the plasma component are shown for $\lambda = 0$ (full line)
and $\lambda = 1$ (dashed line).
}
\end{figure}

In Fig.~\ref{f5} the density perturbations are shown for a two-component
system with an equal amount of perfect fluid and relativistic plasma.
There is little difference from the case considered in Ref.~\cite{Rebhan92a},
where the relativistic plasma was collisionless. The main effect
is again a diminished phase velocity, which is exhibited in
the magnified Fig.~\ref{f6}. There is no longer a simple decay-law for
the density perturbations in the plasma component, because it is
strongly influenced by the comparatively large over- and underdensities
created by the acoustic waves propagating in the perfect fluid component.


\begin{figure}                               
\centerline{ \epsfbox{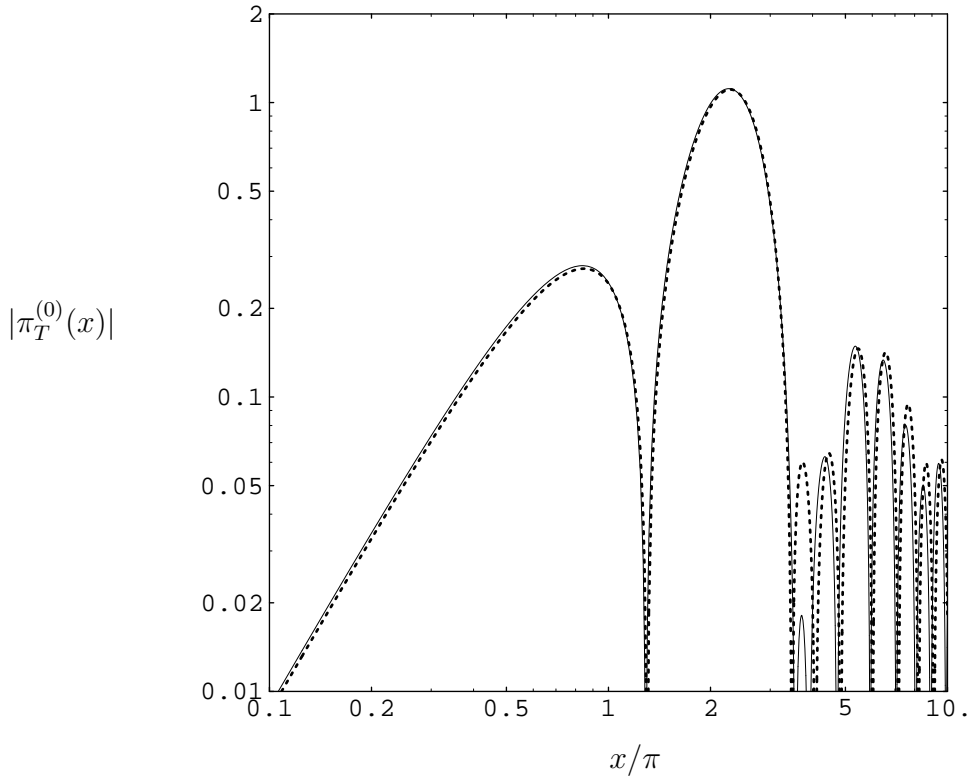} }
\vspace{-4cm}
\unitlength1cm \begin{picture}(16,3)
\put(0,4.5){$|\pi^{(0)}_T(x)|$}
\put(8,-1.3){$x/\pi$}
\end{picture} \\[40pt]
\caption{\label{f7}
The anisotropic pressure is plotted for $\alpha = 1/2$. The full line
shows the behaviour for a plasma with $\lambda = 0$, whereas
the dotted line shows the solution for $\lambda =1$.
}
\end{figure}

In Fig.~\ref{f7}, the anisotropic pressure associated with the
scalar perturbations in the two-component case is given, compared
with the collisionless version. Except for the third peak, there
is remarkably little difference between $\lambda=0$ and $\lambda=1$,
although one might have expected that the anisotropic pressure
would be the most sensitive quantity to self-interactions in
the plasma, since a collision-dominated perfect fluid forbids
anisotropic pressure completely.

\subsection{Vector perturbations}

As we have mentioned, the presence of a relativistic plasma opens the
possibility of having regular vector perturbations. In a perfect-fluid
alone, the velocity amplitude is constant in the radiation-dominated
epoch by virtue of the Kelvin-Helmholtz circulation theorem \cite{HKth}, which
leads to a singular behaviour of the frame dragging potential.
Adding in a plasma component with initially compensating vorticity, however,
allows nontrivial regular solutions. These solutions were first
found within the thermal-field-theoretical treatment and it was
pointed out in Ref.~\cite{Rebhan92b} that they might have interesting
applications in the open issue of primordial magnetic fields.

\begin{figure}                               
\centerline{ \epsfbox{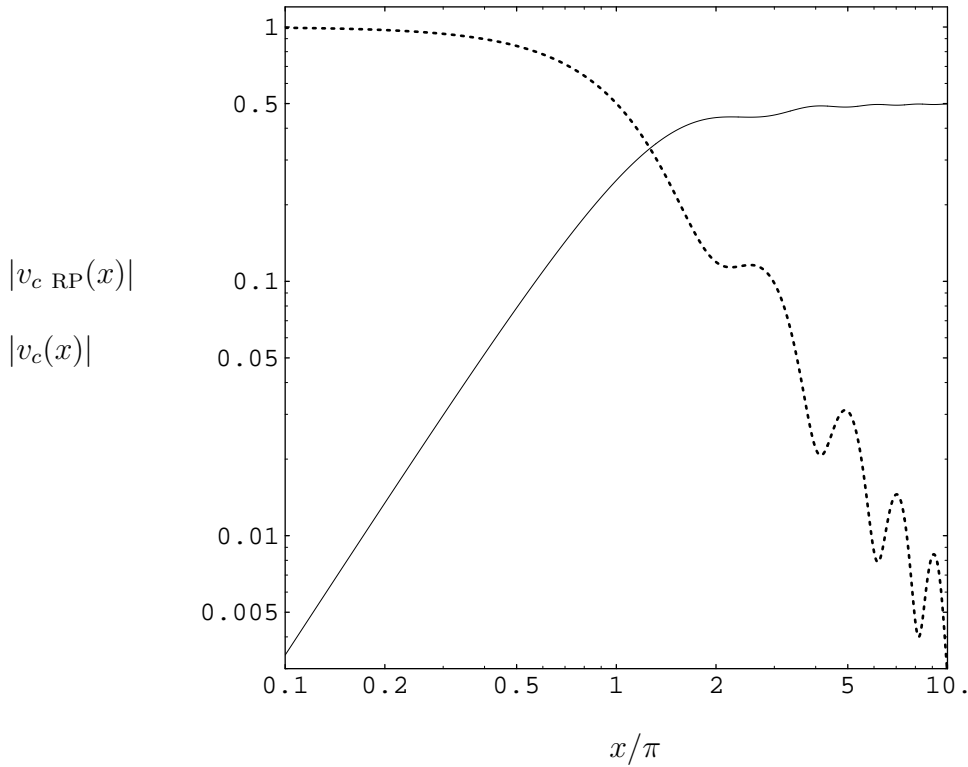} }
\vspace{-4cm}
\unitlength1cm \begin{picture}(16,3)
\put(0,5){$|v_{c\ {\rm RP}}(x)|$}
\put(0,4){$|v_c (x)|$}
\put(8,-1.3){$x/\pi$}
\end{picture} \\[40pt]
\caption{\label{f8}
In a universe containing a mixture ($\alpha = 1/2$)
of a perfect fluid and an ultrarelativistic
plasma rotational perturbations $|v_c (x)|$
may survive in the subhorizon
region (full line). The dotted line shows the rotational perturbation
of the plasma component $|v_{c\ {\rm RP}}(x)|$ with $\lambda = 1$.
}
\end{figure}

\begin{figure}                               
\centerline{ \epsfbox{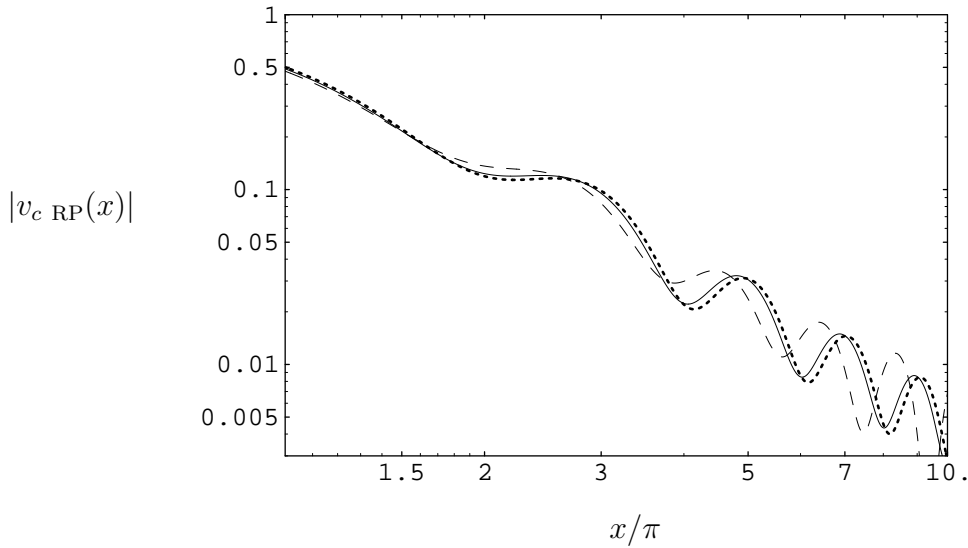} }
\vspace{-4cm}
\unitlength1cm \begin{picture}(16,3)
\put(0,4.5){$|v_{c\ {\rm RP}}(x)|$}
\put(8,0.2){$x/\pi$}
\end{picture} \\[5pt]
\caption{\label{f9}
For the same mixture as in Fig.~8, the rotational perturbation
$|v_{c\ {\rm RP}}(x)|$ is plotted for $\lambda = 0$ (full line)
and for $\lambda = 1$. In the latter case the dashed line
shows the solution through order $\lambda^{3/2}$; the
Pad\'e improved solution is shown by the dotted line.
}
\end{figure}

In Fig.~\ref{f8}, such a solution exhibiting the generation of
a net vorticity which approaches a constant velocity amplitude
is given for $\lambda=1$. In this case there is even only a
relatively small deviation from the collisionless scenario without
the Pad\'e improvement (Fig.~\ref{f9}).
With it, however, the difference becomes
even rather tiny. This is also somewhat unexpected, since the very
existence of these solutions hinges on having a plasma component
that is approximately collisionless. The effect of self-interactions
are found to give only a small increase of the period of the wiggles
in the vorticity of the plasma component while it dies from directional
dispersions.

\begin{figure}                               
\centerline{ \epsfbox{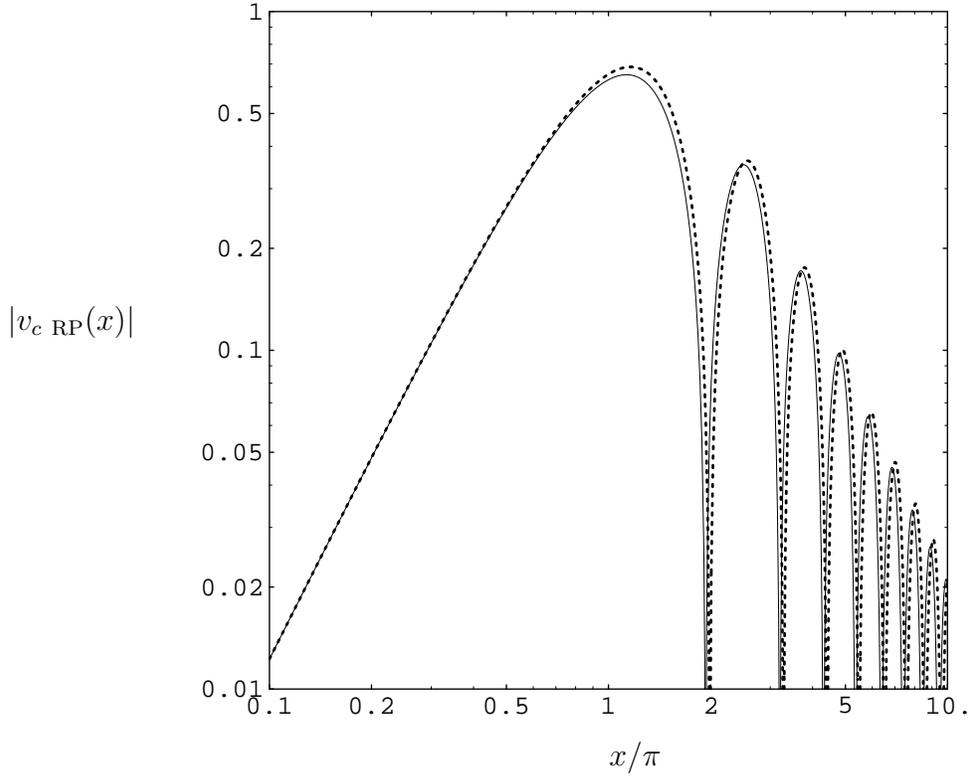} }
\vspace{-4cm}
\unitlength1cm \begin{picture}(16,3)
\put(0,4.5){$|v_{c\ {\rm RP}}(x)|$}
\put(8,-1.3){$x/\pi$}
\end{picture} \\[40pt]
\caption{\label{f10}
Without perfect fluid all rotational perturbations decay on subhorizon scales.
The full line shows the behaviour of collisionless plasmas
($\lambda = 0$), whereas the dotted line shows the Pad\'e improved
solution for $\lambda = 1$.
}
\end{figure}

\begin{figure}
\vspace{-1cm}                              
\centerline{ \epsfbox{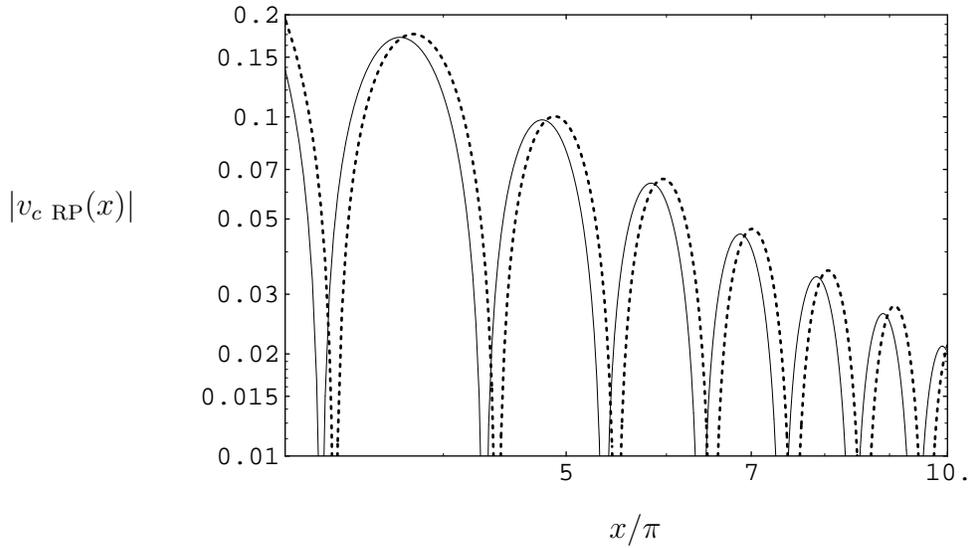} }
\vspace{-4cm}
\unitlength1cm \begin{picture}(16,3)
\put(0,4.5){$|v_{c\ {\rm RP}}(x)|$}
\put(8,0.2){$x/\pi$}
\end{picture} \\[5pt]
\caption{\label{f11}
The same as in Fig.~10, but just the subhorizon region.
}
\end{figure}

Without a perfect-fluid component, one can have a regular solution
which has a growing velocity amplitude on superhorizon scales and
which decays from directional dispersion after horizon crossing,
which is shown in Fig.~\ref{f10}. In Fig.~\ref{f11}, a magnified
picture of the subhorizon behaviour is given, which shows both
a small decrease of the phase velocity of the oscillations and
a small reduction of the damping. While this is similar to the
subhorizon behaviour encountered in the scalar case, it could
hardly be anticipated by comparison with the perfect fluid case,
because in the limit of tight coupling these perturbations are
forbidden entirely.

\subsection{Tensor perturbations}

\begin{figure}                               
\centerline{ \epsfbox{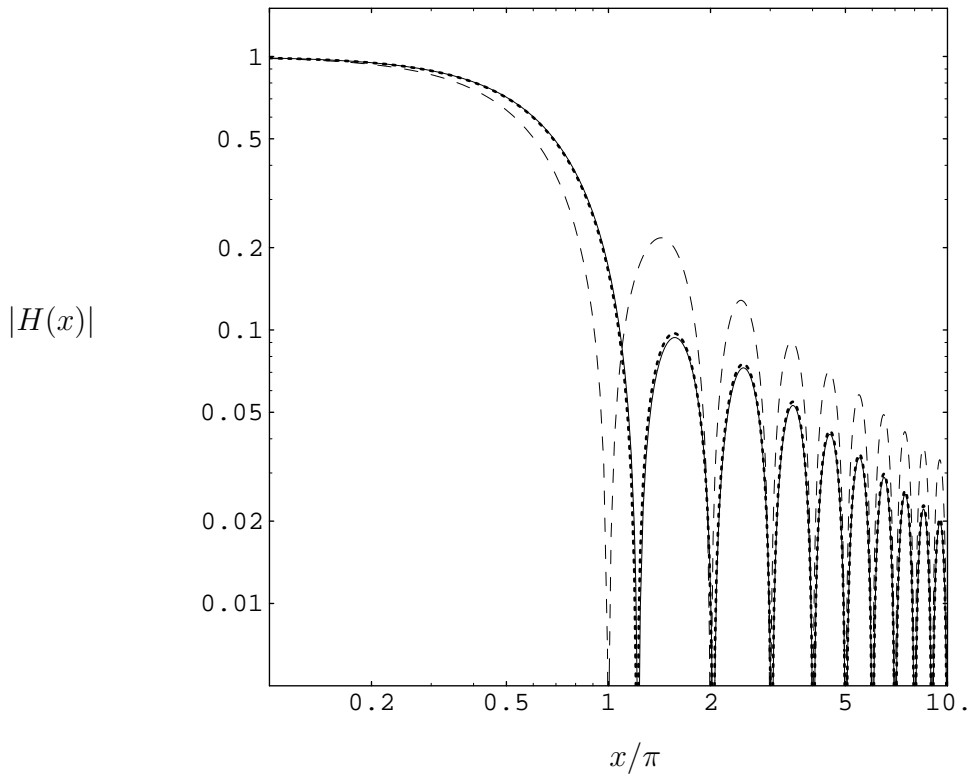} }
\nopagebreak
\vspace{-4cm}
\nopagebreak
\unitlength1cm \begin{picture}(16,3)
\put(0,4.5){$|H(x)|$}
\put(8,-1.3){$x/\pi$}
\end{picture} \\[40pt]
\caption{\label{f12}
The amplitude of a gravitational wave $|H(x)|$ for a
a perfect fluid (dashed line) and an
ultrarelativistic plasma.
The full line solution corresponds to $\lambda = 0$ and the dotted
line to the Pad\'e improved solution for $\lambda = 1$.
}
\end{figure}

Tensor perturbations correspond to primordial gravitational waves.
Their large-time behaviour is expected to be rather independent
of the medium, since it is dictated by energy conservation.
Indeed, there is only some difference in the behaviour at
the time of horizon crossing which implies that a relativistic
plasma requires stronger initial tensor perturbations in order
to have equal amplitude in the gravitational waves at late times.
There is, however, extremely little difference in the behaviour
of the solutions for the plasma case for $\lambda=0$ and $\lambda=1$,
see Fig.~\ref{f12}. The self-interacting case is closer to the
perfect-fluid case, but only very little so.

\begin{figure}                               
\centerline{ \epsfbox{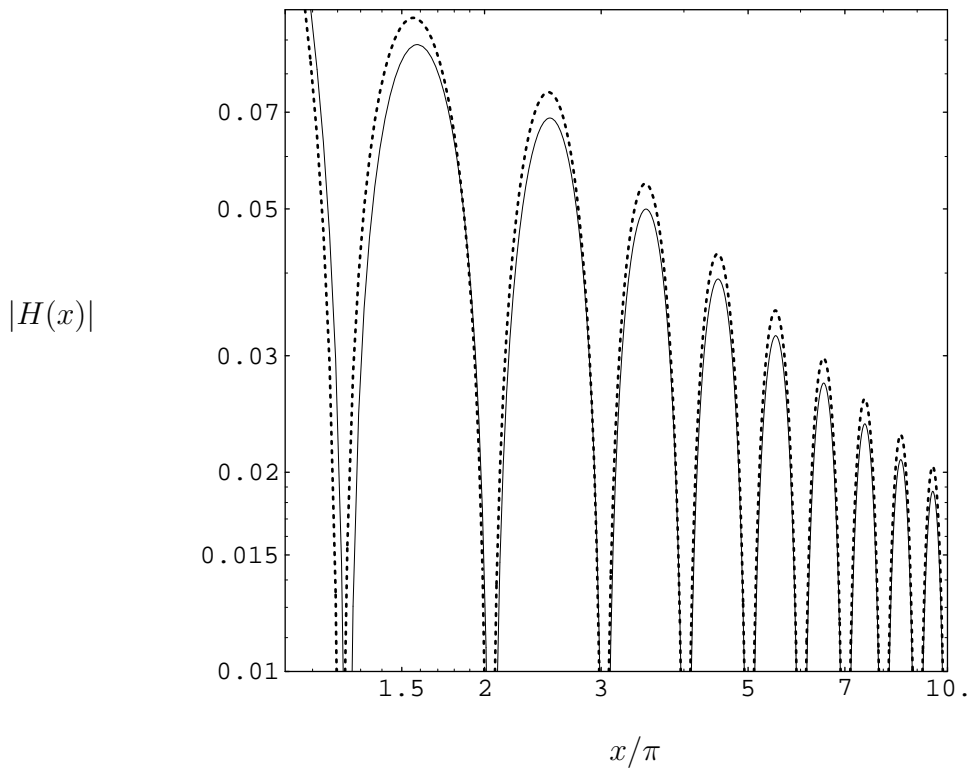} }
\vspace{-4cm}
\unitlength1cm \begin{picture}(16,3)
\put(0,4.5){$|H(x)|$}
\put(8,-1.3){$x/\pi$}
\end{picture} \\[40pt]
\caption{\label{f13}
As in Fig.~12, but only for $\alpha = 1$ with $\lambda = 1$.
The full line shows the solution through order $\lambda^{3/2}$
and the dotted line the Pad\'e improved solution.
}
\end{figure}

As Fig.~\ref{f13} shows, the perturbative result is moreover
rather insensitive
to the Pad\'e-improvement.

\section{Conclusion}

We have studied the effects of weak self-interactions in an
ultrarelativistic plasma on cosmological perturbations through
order $\lambda^{3/2}$ in $\lambda\phi^4$-theory.
At this order it turns out that perturbation theory requires
a resummation of an infinite set of higher-order diagrams. This
is natural to include in the thermal-field-theory approach to
cosmological perturbations, which in the collisionless case is
equivalent to the usual approach based on classical kinetic theory,
but now leaves the latter clearly behind.

While it still turned out to be possible to exactly solve the
perturbation equations by means of a power series ansatz, we have
found that the relatively large coefficients of the order-$\lambda^{3/2}$
corrections make the results reliable only for rather small values
of $\lambda$, and even then there is a breakdown of perturbation
theory in the asymptotic late-time behaviour. The latter comes
from increasingly singular contributions to the gravitational
polarization tensor for light-like momenta and could be cured
by a further resummation procedure. However, the effects of
the latter turned out to be
well approximated by a (2,1)-Pad\'e-improvement of the perturbative
results, which also drastically improves the apparent convergence
of the results for smaller times.

The concrete results obtained showed a tendency toward perfect-fluid
behaviour, but with $\lambda=1$ all of them are still (perhaps surprisingly)
close to the collisionless case. The main effects turned out to
be a small increase of the critical mixing factor of a perfect-fluid
component with a relativistic plasma above which one can have
singular solutions exhibiting superhorizon oscillations. Concerning
the regular solutions, we have found a decrease of the phase velocity
of scalar perturbations and a reduction of its damping. In the
case of vector perturbations (corresponding to
large-scale vorticity), which are only possible in the presence
of a relativistic plasma, similar effects were found, but quantitatively
much smaller. Very little effects from self-interactions were finally
observed in the case of tensor perturbations, which correspond to
primordial gravitational waves.

{}From this one may conclude that a description of a primordial
plasma built from weakly interacting elementary particles
through perfect-fluid models is in general applicable only for scales
far below the Hubble radius. At the scale of the horizon and beyond,
self-interactions can be treated perturbatively, at least in the
model considered here, and there can be significant differences from
the perfect-fluid behaviour, in particular in the case of
rotational perturbations.

\acknowledgments

This work was supported partially
by the Austrian ``Fonds zur F\"orderung der
wissenschaftlichen Forschung (FWF)'' under projects no.\ P9005-PHY
and P10063-PHY, and by the EEC Programme ``Human Capital
and Mobility'',
contract CHRX-CT93-0357 (DG 12 COMA).

\appendix
\section{Gravitational polarisation tensor components and its
Fourier transforms}

The Fourier transform of the combinations of the gravitational
polarisation tensor components \cite{NRS2}
\begin{eqnarray}\label{Ar}
A &=& - \omega\,{\rm artanh}{1\over \omega} + \frac54\nonumber\\
&-&{5\lambda\over 8\pi^2}\left[ 2\left(\omega\,{\rm artanh}{1\over \omega}
\right)^2
-\omega\,{\rm artanh}{1\over \omega}-{\omega^2\over {\omega^2-1}} \right]
\nonumber\\
&-&{5\lambda^{3/2}\over 8\pi^3} \biggl[
3\left({\omega^2-1}-\omega\sqrt{\omega^2-1}\right)\left(\omega\,{\rm artanh}
{1\over \omega}\right)^2 \nonumber\\
&&+6 \left( \omega\sqrt{\omega^2-1}-\omega^2-{\omega\over \sqrt{\omega^2-1}}
\right)\omega
\,{\rm artanh}{1\over \omega}\nonumber\\
&&+{\omega\over ({\omega^2-1})^{3/2}}+3{\omega^2\over {\omega^2-1}}+
6{\omega\over \sqrt{\omega^2-1}}-3\omega\sqrt{\omega^2-1}+3\omega^2
\biggr]  , \nonumber \\
\label{Br}
B &=& -1 
+ {5\lambda\over 4\pi^2}\left[-({\omega^2-1})
\left(\omega\,{\rm artanh}{1\over \omega}\right)^2+(2\omega^2-1)
\omega\,{\rm artanh}{1\over \omega}-\omega^2\right]\nonumber\\
&+&{15\lambda^{3/2}\over 8\pi^3} \biggl[ \left\{ \omega({\omega^2-1})^{3/2}-
({\omega^2-1})^2 \right\}
\left(\omega\,{\rm artanh}{1\over \omega}\right)^2 \nonumber\\
&&+2\left\{ ({\omega^2-1})^2-\omega({\omega^2-1})^{3/2}+\omega\sqrt{\omega^2-1}
-\omega^2 \right\}\omega\,{\rm artanh}{1\over \omega} \nonumber\\
&&-{\omega\over
\sqrt{\omega^2-1}}-2\omega\sqrt{\omega^2-1}+\omega({\omega^2-1})
^{3/2}-\omega^4+4\omega^2 \biggr],
\nonumber \\
\label{Cr}
C &=& - {5\lambda\over 8\pi^2}\left[
3 ({\omega^2-1})^2 \left(
\omega\,{\rm artanh}{1\over \omega}\right)^2 -2({\omega^2-1})(3\omega^2-2)
\omega\,{\rm artanh}{1\over \omega}+3\omega^4-4\omega^2\right]\nonumber\\
&-&{15\lambda^{3/2}\over 16\pi^3} \biggl[
3({\omega^2-1})^2\left({\omega^2-1}-\omega\sqrt{\omega^2-1}\right)
\left(\omega\,{\rm artanh}{1\over \omega}\right)^2 \nonumber\\
&&-2({\omega^2-1}) \left\{ 3({\omega^2-1})^2-3\omega({\omega^2-1})^{3/2}
+3\omega\sqrt{\omega^2-1}-3\omega^2+1 \right\}\omega\,{\rm artanh}
{1\over \omega}
\nonumber\\
&&-3\omega({\omega^2-1})^{5/2}+6\omega({\omega^2-1})^{3/2}+
\omega\sqrt{\omega^2-1}
+3\omega^6-15\omega^4+16\omega^2 \biggr],
\end{eqnarray}
with powers of $\omega$, as required in
(\ref{mv},\ref{vect},\ref{tk}),
boils down to the evaluation of the following integrals
\begin{eqnarray}
&&\int_{-1}^1 {d\omega \over \pi}\cos(\omega x)
(1-\omega^2)^{\beta-\frac{1}{2} } = \left( \frac{2}{x} \right)^\beta
\frac{\Gamma (\beta + \frac{1}{2} ) }{\sqrt{\pi} } J_\beta (x)  ,\nonumber \\
&&\frac{1}{2} \int_{-1}^1 d\omega \sin (\omega x)
\ln{1+\omega\over 1-\omega}  =
 {1\over x}\left[ \sin(x) {\rm Si}(2x) + \cos (x)
\left( {\rm Ci}(2x) - \gamma - \ln (2x)\right) \right] \nonumber \\
&& =  2\sum_{m=0}^{\infty} {(-1)^m x^{2m+1}\over (2m+2)!} \sum_{j=0}^m
{1\over 2j+1} =: \kappa(x) ,  \nonumber   \\
&&\int_{-1}^1 {d\omega \over 2 \pi}
\cos(\omega x) {\omega\over\sqrt{1-\omega^2}}
\ln{1+\omega\over 1-\omega}   =:  \nu(x) , \nonumber \\
&&- 12 \int_{-1}^1 {d\omega \over 2 \pi}
\cos(\omega x) \sqrt{1-\omega^2}
\left({\omega\over 2} \ln{1+\omega \over 1-\omega}\right)^2
-  {3\pi^2\over2} \left(J_1\over x\right)^{\prime\prime}
=: \xi(x)
\end{eqnarray}
where additional powers of $\omega$ in the integrands can  be
obtained by suitably differentiating those functions with respect
to $x$. Let us first consider a typical integral appearing in the
power-series expansion  in $x$ of $\nu(x)$,
$$ S_m =  \int_{-1}^{1} d\omega \frac{\omega^{2 m+1}}{\sqrt{1-\omega^2 } }
\ln \frac{1+\omega }{1-\omega} .$$
By integration by parts with respect to the root, it can be shown to
satisfy
the recursion $S_m= 2 m ( S_{m-1} -S_{m} ) +
 2 \sqrt{\pi} \Gamma ( \frac12  + m) / m! $ which may be solved for
 $S_m$,
$$S_m = 2 \sqrt{\pi} \sum_{l=0}^{n} \frac{\Gamma (\frac12 +l)  }{l!
(1 + 2 m - 2 l ) },$$
that finally leads to the expression
\begin{eqnarray}
\nu(x)
&=& \sum_{m=0}^\infty {(-1)^m x^{2m} \over (2m)!} \sum_{l=0}^m
{\Gamma(l+\frac12) \over \sqrt{\pi} l! (1+2m-2l)}
.\end{eqnarray}
The square of the logarithm in the
more complicated power-series coefficients of $\xi (x) $,
$$T_m =   \int_{-1}^{1} d\omega  \omega^{2 m +2} \sqrt{1-\omega^2 }
\ln^2 \frac{1+\omega }{1-\omega} ,$$
can be removed by integrating by parts in the same manner as above
with the remaining
integral being of the type of $S_m$ with the root in the numerator.
$T_m$ satisfies the recursion $ 3 T_m =  (2 m +1 ) ( T_{m-1} - T_m )
+ 4 ( S_m - S_{m+2} ) $ which can be solved using the initial
integral $T_0 = \pi^3/8 - 2 \pi /3  $. Splitting off the
contributions $\propto \pi^3$ which can be reexpressed in terms
of $J_1(x)$ finally yields
\begin{equation}
\xi(x)
= 6 \sum_{m=0}^\infty {(-1)^m x^{2m} \over (2m)!} {\Gamma(m+\frac32)\over
\Gamma(\frac12) \Gamma(m+3)}
 \left(1 -
\sum_{l=0}^m {\Gamma(l+1)\over 4 \Gamma(l+\frac 32)} \sum_{j=0}^l
{\Gamma(j+\frac12)\over j! (l - j + \frac32)}\right) \ .
\end{equation}

\section{Recursion relations}

In order to solve the integro-differential equations for the
metric potentials a power series ansatz is appropriate. The integral
kernels are expanded as
\begin{equation}
K^{(a)}(x) = \sum_{n=0}^{\infty} (-1)^n K^{(a)}_n {x^{2n}\over (2n)!} \ .
\end{equation}
They read:
\begin{eqnarray}
K^{(0)}_n &=& {1\over 2n+1} + \frac 58 {\lambda\over \pi^2}
\left( {2\over n+1} \sum_{j=0}^n {1\over 2j+1} - 1 - {1\over 2n+1} \right) +
\nonumber \\
& & + \frac{15}8 \left({\lambda\over\pi^2}\right)^{\frac32}
\left[1 + {(2n-1)!!\over(2n-2)!!}\left({1\over 3} + {1\over n} +
{1\over 2n(2n+2)}\right) - {6\over 2n+3} - \right. \nonumber \\
& & \qquad - {1\over (n+1)(n+2)} \sum_{j=0}^n {1\over 2j+1}
+ {1\over (n+2)(2n+3)} - \nonumber \\
& & \qquad -{4\over \sqrt{\pi}}\sum_{j=0}^n{\Gamma(j+\frac12)\over j!(2n-2j+1)}
+ {2\over \sqrt{\pi}}\sum_{j=0}^{n+1}{\Gamma(j+\frac12)\over j!(2n-2j+3)} -
\nonumber \\
& & \qquad \left. - {\Gamma(n+\frac32)\over\sqrt{\pi}(n+2)!}\left(1-
\sum_{l=0}^n {l!\over 2 \Gamma(l+\frac32)} \sum_{j=0}^l
{\Gamma(j+\frac12)\over j! (2l-2j+3)} \right) \right] \, \\
K^{(1)}_n &=& - {1\over (2n+1)(2n+3)} + \frac 54 {\lambda\over \pi^2}
{n\over (2n+1)(2n+3)} - \nonumber \\
& & - \frac 58 \left({\lambda\over\pi^2}\right)^{\frac32}
\left({(2n-1)!!\over (2n)!!} - {1\over 2n+1} +
{8n \over (2n+1) (2n+3)}\right) \ , \\
K^{(2)}_n &=& -{8\over (2n+1)(2n+3)(2n+5)} + 5 {\lambda\over \pi^2}
{1\over (2n+1)(2n+3)}\left(1 - {3\over2n+5}\right) - \nonumber \\
& & - 5 \left({\lambda\over\pi^2}\right)^{\frac32}
\left[{(2n-1)!!\over (2n +2)!!}+
{3\over(2n+1)(2n+3)}\left(1-{4\over2n+5}\right)\right] \ .
\end{eqnarray}
In this Appendix we restrict our attention to regular even solutions, as the
kernels are even functions of $x$. This means that we assume
$\gamma_{2n+1}^{(a)} = 0$ for all $n$ and $a$. Regular odd and singular
solutions may be obtained in a similar way \cite{Rebhan94}.

\subsection{Scalar Perturbations}

With
\begin{equation}
\Phi_N = \sum_{n=0}^{\infty} (-1)^n \phi_n {x^{2n}\over (2n)!}
\end{equation}
Eq.~(\ref{phiN}) leads to the recursion relation
\begin{eqnarray}
\left[6n + 9 - {9 \alpha\over n+1}
\left(K_1^{(0)} - 3 K_2^{(0)}\right) \right] \phi_n &=&
- 4 \sum_{l=1}^{n-1} (l+2)\phi_{l} + \nonumber \\
&+&
{9 \alpha\over n+1} \sum_{l=1}^{n-1} \left(K_{l+1}^{(0)} - 3 K_{l+2}^{(0)}
\right) \phi_{n-l} +
\nonumber \\
&+& {9 \alpha\over n+1} \sum_{l=0}^{\infty} (-1)^l \gamma_{2l}^{(0)}
\left(K_{n+1+l}^{(0)} - 3 K_{n+2+l}^{(0)}\right) + \nonumber \\
&+& \phi_0 - 9 \alpha \sum_{l=0}^{\infty} (-1)^l \gamma_{2l}^{(0)}
\left(K_{1+l}^{(0)} - 3 K_{2+l}^{(0)}\right)\ ,
\end{eqnarray}
for $n\ge 1$. The regularity at $x=0$ demands
\begin{equation}
\phi_0 = C_1 + \frac 23 \sum_{l=0}^{\infty} \gamma_{2l} (-1)^l (2n -4) \ .
\end{equation}
Accordingly, the metric perturbation
\begin{equation}
\Pi = \sum_{n=0}^{\infty} (-1)^n \pi_n {x^{2n}\over (2n)!}
\end{equation}
is determined from Eqs.~(\ref{pit}) and (\ref{pitk}). Its recursion reads:
\begin{eqnarray}
\pi_n &=& {3\alpha \over (n+1)(2n+1)}\left[ \sum_{l=1}^n \left(
K^{(0)}_l - 3 K^{(0)}_{l+1} \right) \phi_{n-l+1} + \right. \nonumber \\
& & \left. + \sum_{l=0}^{\infty} \gamma_{2l} (-1)^l \left(
K^{(0)}_{n+l+1} - 3 K^{(0)}_{n+l+2} \right) \right]
\end{eqnarray}
for $n\ge 0$. The relativistic plasma component perturbation
\begin{equation}
\Phi_{\rm RP} = \sum_{n=0}^{\infty} (-1)^n q_n {x^{2n}\over (2n)!}
\end{equation}
follows from (\ref{phiRP}) and reads
\begin{eqnarray}
q_n &=& {3\alpha \over (n+1)(2n+1)} \left[ -\phi_{n+1} - \pi_{n+1} +
2 \sum_{l=0}^n \left(K^{(0)}_l + {3\over 2n+3} K^{(0)}_{l+1} \right)
\phi_{n-l+1} + \right. \nonumber \\
& & \left. + 2 \sum_{l=0}^{\infty} \gamma_{2l} (-1)^l \left(
K^{(0)}_{n+l+1} + {3\over 2n+3} K^{(0)}_{n+l+2} \right) \right]
\end{eqnarray}
for $n\ge 0$. From regularity at $x=0$
\begin{equation}
C_1 = \sum_{l=0}^{\infty} \gamma_{2l} (-1)^l \left(-\frac23 (2l -4)+
9 \alpha K^{(0)}_{l+2} - 3(\alpha -2) K^{(0)}_{l+1} + 2K^{(0)}_l \right)
\end{equation}
follows.

\subsection{Vector Perturbations}

The vector metric potential
\begin{equation}
\Psi = \sum_{n=0}^{\infty} (-1)^n \psi_n {x^{2n}\over (2n)!}
\end{equation}
is determined by the Eqs.~(\ref{vector}) and (\ref{vk}). The corresponding
recursion is
\begin{eqnarray}
\left[(n+1)(2n+1) - 12\alpha K_1^{(1)}\right] \psi_n &=&
12 \alpha \sum_{l=1}^{n} K_{l+1}^{(1)} \psi_{n-l} +
\nonumber \\
&+& 12 \alpha \sum_{l=0}^{\infty} (-1)^l \gamma_{2l}^{(1)} K_{n+1+l}^{(1)} \ ,
\end{eqnarray}
for $n\ge 0$. The relation (\ref{vPF}) couples the vorticity of the perfect
fluid component to the coefficients $\gamma_{2l}^{(1)}$.

\subsection{Tensor Perturbations}

For the metric potential $H$ the ansatz
\begin{equation}
H = \sum_{n=0}^{\infty} (-1)^n h_n {x^{2n}\over (2n)!}
\end{equation}
is made. Regular even solutions follow from Eqs.~(\ref{tensor}) and (\ref{tk})
by
\begin{eqnarray}
\left[ (2n+1)2n - 3\alpha K_0^{(2)} \right] h_n &=&
2n(2n-1) h_{n-1}
+ 3 \alpha \sum_{l=1}^{n-1} K_l^{(2)} h_{n-l} + \nonumber \\
&+& 3 \alpha \sum_{l=0}^{\infty} (-1)^l \gamma_{2l}^{(2)} K_{n+l}^{(2)}
\end{eqnarray}
for $n\ge 1$. The initial value $h_0$ specifies together with the
inhomogeneous terms the regular solution.


\begin{references}
\bibitem{Lifshitz}E. Lifshitz, Zh. Eksp. Teor. Fiz. {\bf 16}, 587 (1946);
         E. Lifshitz and I. Khalatnikov, Adv. Phys. {\bf 12}, 185 (1963).
\bibitem{Kodama}H. Kodama and M. Sasaki, Prog. Theor. Phys. Suppl.
         {\bf 78}, 1 (1984).
\bibitem{Mukhanov}V. F. Mukhanov, H. A. Feldman, and R. H. Brandenberger,
         Phys. Rep. {\bf 215}, 203 (1992); R. Durrer, {\it Gauge Invariant
         Cosmological Perturbation Theory}, Report No. ZU-TH14/92, 1993.
\bibitem{Peebles70}P. J. E. Peebles and J. T. Yu, Astrophys. J. {\bf 162}, 815
 (1970).
\bibitem{McCone70}A. I. McCone, Ph. D. thesis, Univ. of Maryland,
        Tech. Rept. UMD-70-054 (1970).
\bibitem{Peebles73}P. J. E. Peebles, Astrophys. J. {\bf 180}, 1 (1973).
\bibitem{Bond83}J. R. Bond and A. S. Szalay,
         Astrophys. J. {\bf 274}, 443 (1983).
\bibitem{Ehlers}J. Ehlers, in {\it General Relativity and Gravitation}, edited
         by R. K. Sachs (Academic Press, New York, 1971); J. M. Stewart,
         {\it Non-equilibrium Relativistic Kinetic Theory}, (Springer-Verlag,
         New York, 1971).
\bibitem{Stewart72}J. M. Stewart, Astrophys. J. {\bf 176}, 323 (1972).
\bibitem{Zakharov}A. V. Zakharov, Sov. Phys. JETP {\bf 50}, 221 (1979);
        E. T. Vishniac, Astrophys. J. {\bf 257}, 456 (1982).
\bibitem{Kraemmer}U. Kraemmer and A. Rebhan, Phys. Rev. Lett. {\bf 67},
 793 (1991).
\bibitem{Rebhan91}A. Rebhan, Nucl. Phys. {\bf B351}, 706 (1991).
\bibitem{ABFT}J. Frenkel and J. C.
         Taylor, Z. Phys. C {\bf49}, 515 (1991);
	F. T. Brandt, J. Frenkel and J. C.
         Taylor, Nucl. Phys. {\bf B374}, 169 (1992);
	A. P. de Almeida, F. T.  Brandt and J. Frenkel,
         Phys. Rev. D {\bf 49}, 4196 (1994);
	J. Frenkel, E. A. Gaffney and J. C. Taylor,
         Nucl. Phys. {\bf B439}, 131 (1995).
\bibitem{Rebhan92a}A. Rebhan, Nucl. Phys. {\bf B368}, 479 (1992);
\bibitem{Schwarz}D. J. Schwarz, Int. J. Mod. Phys. D {\bf 3}, 265 (1994).
\bibitem{Rebhan94}A. K. Rebhan and D. J. Schwarz,  Phys. Rev. D
         {\bf 50}, 2541 (1994).
\bibitem{Kasai}M. Kasai and K. Tomita, Phys. Rev. D {\bf 33}, 1576 (1986).
\bibitem{NRS1}H. Nachbagauer, A. K. Rebhan, and D. J. Schwarz, Phys. Rev. D
         {\bf 51}, R2504 (1995).
\bibitem{NRS2}H. Nachbagauer, A. K. Rebhan, and D. J. Schwarz, {\it
 The gravitational polarization tensor of thermal $\lambda \phi^4$
 theory}, preprint hep-th/9507099 (1995).
\bibitem{Bardeen}J. M. Bardeen, Phys. Rev. D {\bf 22}, 1882 (1980).
\bibitem{Kapusta}J. I. Kapusta, {\em Finite-temperature field theory},
         (Cambridge University Press, Cambridge, 1989);
M. Le Bellac, {\em Thermal Field Theory} to appear in
Cambridge University Press.
\bibitem{BP}E. Braaten and R. D. Pisarski, Nucl. Phys. {\bf B355}, 1 (1991).
\bibitem{Ellis}M. Bruni, P. K. S. Dunsby, and G. F. R. Ellis,
 Astrophy. J. {\bf 395}, 34 (1992).
\bibitem{Grishchuk}L. P. Grishchuk, Phys. Rev. D {\bf 50}, 7154 (1994).
\bibitem{Deruelle}N. Deruelle and V. F. Mukhanov, {\em On matching conditions
        for cosmological perturbations}, gr-qc/9503050 (1995).
\bibitem{Schaefer}R. K. Schaefer and A. A. de Laix, {\em Gauge invariant
        density and temperature perturbations in the quasi-Newtonian
        formulation}, astro-ph/9507003 (1995).
\bibitem{Rebhan92b}A. Rebhan, Astrophys. J. {\bf392}, 385 (1992).
\bibitem{Wolfram}S. Wolfram, {\em Mathematica} 2nd Ed., (Addison-Wesley,
        Redwood City, 1991).
\bibitem{Boerner}G. B\"orner, {\em The Early Universe: Facts and Fiction}
        (Springer, Berlin, 1993).
\bibitem{DolanJackiw}L. Dolan and R. Jackiw,
        Phys. Rev. D {\bf 9}, 3320 (1974).
\bibitem{Pade}G.~A.~Baker, Jr., {\it Essentials of Pad\'e Approximants},
        (Academic Press, New York, 1975).
\bibitem{Dav}B. Davies, {\it Integral Transforms and Their
        Applications}, (Springer-Verlag, New York, 1978).
\bibitem{HKth}B. J. T. Jones, Rev. Mod. Phys. {\bf 48}, 107 (1976).
\end{references}
\end{document}